\documentclass[useAMS,usenatbib]{mn2e}

\newcommand{\kms}{ km s$^{-1}$}
\newcommand{\ang}{\mbox{\AA} }
\usepackage{graphicx}
\usepackage{times}
\usepackage{colordvi}
\usepackage{epsfig}
\usepackage{lscape}
\usepackage{rotating}
\usepackage{rotate}
\usepackage{amsfonts}
\usepackage{array}

\begin{document}

\title[Metals in  sub-Damped Lyman-$\alpha$ Absorbers at $1.7<z<2.4$]{Element Abundances at High-redshift: Magellan MIKE Observations of sub-Damped Lyman-$\alpha$ Absorbers at
$1.7<z<2.4$}

\author[D. Som et al.]{Debopam Som$^{1}$, 
Varsha P. Kulkarni$^{1}$, 
Joseph Meiring$^{2}$, 
Donald G. York$^{3,4}$,
Celine P\'eroux$^{5}$,
\newauthor  Pushpa Khare$^{6}$, and James T. Lauroesch$^{7}$\\
$^{1}$Department of Physics and Astronomy, University of South Carolina, Columbia, SC 29208, USA \\
$^{2}$Department of Astronomy, University of Massachusetts, Amherst, MA 01003, USA \\ 
$^{3}$Department of Astronomy and Astrophysics, University of Chicago, Chicago, IL 60637, USA \\
$^{4}$Enrico Fermi Institute, University of Chicago, Chicago, IL 60637, USA \\
$^{5}$Aix-Marseille Universit\'e, CNRS, LAM(Laboratoire d'Astrophysique de Marseille) UMR 7326, 13388, Marseille, France \\
$^{6}$CSIR Emeritus Scientist, IUCAA, Pune, India \\
$^{7}$Department of Physics and Astronomy, University of Louisville, Louisville, Ky 40292 USA\\}

\date{Accepted: 2013 July 24, Received: 2013 July 20; in original form: 2012 November 09}

\pagerange{\pageref{firstpage}--\pageref{lastpage}} \pubyear{2013}

\maketitle

\label{firstpage}

\begin{abstract}
We present chemical abundance measurements from high-resolution observations of 5 sub-damped Lyman-$\alpha$ absorbers at $1.7 < z < 2.4$ observed with
the Magellan Inamori Kyocera Echelle (MIKE) spectrograph on the 6.5-m Magellan II Clay telescope. Lines of Zn II, Mg I, Mg II, Al II, Al III, S II, Si II,
Si IV, C II, C II*, C IV, Ni II, Mn II and Fe II were detected and column densities were determined. The metallicity of the absorbing gas, inferred from
the nearly undepleted element Zn, is in the range of $<-0.95$ to $+0.25$ dex for the five absorbers in our sample, with three of the systems being near-solar
or super-solar. We also investigate the effect of ionisation on the observed abundances using photoionisation modelling. Combining our data with other
sub-DLA and DLA data from the literature, we report the most complete existing determination of the metallicity vs. redshift relation for sub-DLAs and DLAs.
We confirm the suggestion from previous investigations that sub-DLAs are, on average, more metal-rich than DLAs and evolve faster. We also discuss relative
abundances and abundance ratios in these absorbers. The more metal-rich systems show significant dust depletion levels, as suggested by the ratios [Zn/Cr]
and [Zn/Fe]. For the majority of the systems in our sample, the [Mn/Fe] vs. [Zn/H] trend is consistent with that seen previously for lower-redshift sub-DLAs.
We also measure the velocity width values for the sub-DLAs in our sample from unsaturated absorption lines of Fe II $\lambda\lambda\lambda$ 2344, 2374,
2600$\ang$, and examine where these systems lie in a plot of metallicity vs. velocity dispersion. Finally, we examine cooling rate vs. H I column density
in these sub-DLAs, and compare this with the data from DLAs and the Milky Way ISM. We find that most of the systems in our sample show higher cooling rate
values compared to those seen in the DLAs.

\end{abstract}


\section{Introduction}
Many open questions remain about the processes of galaxy formation and evolution. Heavy element abundance measurements in galaxies reveal important information
about the ongoing processes of star formation and death, and the overall chemical enrichment of these galaxies. Studying the chemical composition of high-redshift
galaxies through emission lines often leads to a bias toward the most actively star-forming galaxies. Moreover, it is difficult to determine abundances accurately
from emission line indices. Quasar Absorption Line Systems (QSOALS) provide a way to study the interstellar medium (ISM) of high redshift galaxies independent
of the morphology and luminosity of galaxies. In addition, absorption lines in QSO spectra allow us to study the diffuse intergalactic medium using UV or X-ray
observations of high-ionization species such as N V, O VI, O VIII, etc. (e.g., \citealt{Sim02, Fang02}).

Quasar absorption line systems with strong Lyman-$\alpha$ lines are often divided into two classes: Damped Lyman-$\alpha$ (DLA, log N$_{\rm H \ I}$ $\ge$ 20.3)
and sub-Damped Lyman-$\alpha$ (sub-DLA, 19 $\la$ log N$_{\rm H \ I}$ $<$ 20.3, \citealt{Per01}). DLAs and sub-DLAs contain a major fraction of the neutral gas
in the Universe, while the majority of the baryons are thought to lie in the highly ionized and diffuse Lyman-$\alpha$ forest clouds with log N$_{\rm H \ I}$
$\le$ 14 in intergalactic space (e.g., \citealt{Petit93}).The DLA and sub-DLA systems are generally believed to be associated directly with galaxies at all redshifts
at which they are found. Indeed, several DLA host galaxies have been confirmed through deep imaging and follow-up spectroscopy (e.g., \citealt{Chen03, Ghar06}).

\begin{table*}
\center	
\footnotesize
\caption{Summary of Observations.}
\label{targets}
\begin{tabular}{cccccccccccc}
\hline
\hline
QSO		&	RA		&	Dec		&	m$_{V}$	&	$z_{em}$	&	$z_{abs}$	&	log N$_{\rm H \ I}$	&	Exposure Time	&	Epoch		&	N$_{\rm H \ I}$	\\
J2000		&			&			&		&			&			&		cm$^{-2}$ 	&	sec		&			&	Reference	\\
\hline
Q1039-2719	&	10:39:21.83	&	-27:19:16.0	&	17.4	&	2.193		&	2.139		&	19.55$\pm$0.15		&	7100		&  2008 March 16 	&		1	\\
Q1103-2645	&	11:03:25.29	&	-26:45:15.7	&	16.0	&	2.145   	&	1.839		&	19.52$\pm$0.04		&	3600		&  2008 March 16 	&		1	\\
Q1311-0120	&	13:11:19.26	&	-01:20:30.9	&	17.5	&	2.585		&	1.762		&	20.00$\pm$0.08		&	8100		&  2008 March 16 	&		2	\\
Q1551+0908	&	15:51:03.39	&	+09:08:49.3	&	17.9	&	2.739		&	2.320		&	19.70$\pm$0.05		&	6300		&  2010 May 06   	&		1	\\
Q2123-0050	&	21:23:29.47	&	-00:50:53.0	&	16.7	&	2.262		&	2.058		&	19.35$\pm$0.10		&	4800		&  2010 May 06	 	&		1	\\
\hline
\end{tabular}

N$_{\rm H \ I}$ References. -- (1) This Work, (2) \citet{Wol95}

\end{table*}

A number of chemical elements are detected in DLAs and sub-DLAs, e.g., C, N, O, Mg, Si, S, Ca, Ti, Cr, Mn, Fe, Ni, and Zn. Among these elements, Zn is often
adopted as the tracer of gas-phase metallicity as it is relatively undepleted in the Galactic ISM, especially when the fraction of H in molecular form is low,
as is the case in most DLAs. Zn also tracks the Fe abundance in Galactic stars (e.g., \citealt{Nissen04}), and the lines of Zn II $\lambda$$\lambda$ 2026,2062
are relatively weak and typically unsaturated. These lines can also be covered with ground-based spectroscopy over a wide range of redshifts, from 0.65
$\la z \la$ 3.5, which covers a large portion of the history of the universe. Abundances of refractory elements such as Cr and Fe relative to Zn also give us
a measure of the amount of dust depletion \citep{York06}. Abundance ratios such as [Si/Fe], [O/Fe] and [Mn/Fe] shed light on the enrichment from the different
types of supernovae, as the $\alpha$-capture elements Si and O are produced mainly in Type II supernovae while the iron peak elements are produced mainly by
Type Ia supernovae.

The majority of previous studies of element abundances have focused on DLAs because of their high gas content \citep{PW02, Kul05, Mei06}. Most DLAs have been found to be
metal poor, typically far below the solar level and below the model predictions for the mean metallicity at the corresponding redshifts at which they are seen (e.g,
\citealt{Kul05} and references therein). We note that DLAs detected in the spectra of gamma-ray burst (GRB) afterglows are generally found to be more metal rich than
their quasar absorber counterparts (e.g., \citealt{Fyn08,Sav09,Sav12}, and references therein). However, the sample of GRB-DLAs is much smaller than that of the QSO-DLAs.
Also, most of the GRB-DLAs arise in GRB host galaxies that are likely to have high specific star formation rates and may not be typical. Also, it is very likely that the difference
between GRB-DLAs and QSO-DLAs may be caused by differences in the regions of the host galaxies probed by them, with GRB-DLAs probing inner star-forming regions and QSO-DLAs
probing outer regions (e.g., \citealt{Fyn08}).

The sub-DLA quasar absorption systems have until recently been largely ignored, so their contribution to the overall metal budget is not well-known. Our recent Magellan,
MMT and VLT data have increased the sub-DLA Zn sample at $0.7 \le z \le 1.5$ by a factor of $>8$, and several metal-rich sub-DLAs including some super-solar systems have
been discovered \citep{Mei06,Mei07,Mei08,Mei09,Per06a,Per06b,Kul07}. Evidence for the possibility of a non-negligible contribution from sub-DLAs to the metal budget came
from Kulkarni et al. (2007, 2010, and references therein) based on Zn abundance measurements (see also \citealt{Per03a} for a similar early suggestion but based on the
strongly depleted element Fe).

In this work, we present high-resolution spectroscopic observations of 5 sub-DLAs taken with the Magellan Inamori Kyocera Echelle (MIKE) spectrograph on the 6.5m
Clay telescope at the Las Campanas Observatory. This paper is structured in the following way: In $\S$ 2, details of our observations and data reduction techniques
are discussed. $\S$ 3 discusses the methods used to determine column densities of various ions. In $\S$ 4, information on the individual absorbers from our sample
are given. In $\S$ 5, we present the results from the analysis of our data, and finally in $\S$ 6, we discuss conclusions drawn from this work.

\section{Observations and Data Reduction}

The spectra of the quasars presented here were obtained over 2 separate epochs, 2008 March and 2010 May, respectively, with the Magellan Inamori Kyocera Echelle
spectrograph (MIKE) \citep{Bern03} on the 6.5m Magellan Clay telescope at Las Campanas Observatory. MIKE is a double sided spectrograph consisting of both a blue
and a red camera, providing for simultaneous wavelength coverage from $\sim$3340 \ang to $\sim$9400 \ang. The sightlines were observed in multiple exposures of
1800 to 2700 seconds each, to minimize cosmic ray defects. During data acquisition, seeing was typically $<$ 1$\arcsec$, averaging $\sim$ 0.7$\arcsec$. The target
QSOs were observed with the 1$\arcsec$x5$\arcsec$ slit and the spectra were binned 2x3 (spatial by spectral) during readout. The resolving power of the MIKE spectrograph
is $\sim$19,000 and $\sim$25,000 on the red and blue sides respectively with a 1$\arcsec$x5$\arcsec$ slit. Table \ref{targets} gives a summary of the observations.

We reduced the spectra using the MIKE pipeline reduction code in IDL developed by S. Burles, J. X. Prochaska, and R. Bernstein. The MIKE software makes use of the
overscan region to perform bias subtraction and then flat-fields the data. The software then performs sky-subtraction and extracts the spectral orders using the
traces from flat field images. The pipeline calibration code uses Th-Ar comparison lamp exposures, taken before and after each science exposure, to perform wavelengths
calibration. The software also corrects for heliocentric velocities and converts the wavelengths to vacuum values. Each individual echelle order was then extracted
from the IDL structure created by the pipeline software and corresponding orders from multiple exposures were combined in IRAF using rejection parameters to reduce
the effects of cosmic rays. The spectra from these combined orders were then normalized individually using Legendre polynomial functions to fit the continuum. Typically,
these functions were of order five or less.

Our sample consists of 5 sub-DLAs at $z>1.7$, including 3 at $z>2$. We focus on this redshift range, because few abundance measurements exist for sub-DLAs at these
redshifts, especially at $z>2$ (e.g., \citealt{DZ03,DZ09,EL01,Led06,Not08,Pet94}). All of the absorbers in our sample have N$_{\rm H \ I}$ values known previously
either from the Large Bright Quasar Survey or measured from the Ly$\alpha \, \lambda$ 1215.7 line seen in SDSS spectra. However, for the absorbers with the Ly$\alpha
\, \lambda$ 1215.7 line falling within the spectral coverage of our MIKE observations (which is the case for all the systems except the absorber toward Q1311-0120),
we report N$_{\rm H \ I}$ values determined from our high resolution data.

\section{Determination of Column Densities}

Column densities were determined by fitting the normalised absorption profiles using the FITS6P package \citep{Wel91}, which has evolved from the code by \citet{Vm77}.
FITS6P iteratively minimizes the $\chi^{2}$ value between the data and a theoretical Voigt profile that is convolved with the instrumental profile. The Voigt profile
fits to the absorption features seen in our data used multiple components, tailored to the individual system. For the central, core components, the Doppler parameters
($b_{eff}$) and radial velocities were determined from the weaker and less saturated lines, typically the Fe II $\lambda$ 2374 or the Mg I $\lambda$ 2852 line. For the
weaker components at higher radial velocities, the $b_{eff}$ and component velocity values were determined from stronger transitions such as Fe II $\lambda\lambda$ 2344,
2382 and Mg II $\lambda\lambda$ 2796, 2803. A set of $b_{eff}$ and $v$ values were thus determined that provide reasonable fits to all of the lines observed in the system.
The atomic data used in the identification of lines and profile fitting were adopted from \citet{Morton03}.

If a multiplet was observed, the lines were fitted simultaneously. For all of the systems, the Fe II $\lambda$ 2344, 2374, 2382 lines were fitted simultaneously to arrive
at a set of column densities that provide reasonable fits to the spectra. Similarly, the Mg II $\lambda\lambda$ 2796, 2803 lines were also fitted together. At the resolution
of our data, the Zn II $\lambda$ 2026 line is blended with the Mg I $\lambda$ 2026 line. The Mg I contribution to the blend was estimated using the Mg I $\lambda$ 2852 line,
for which f$\lambda$ $\sim$32 times that of the Mg I $\lambda$ 2026 line. The Zn II contribution was determined by fitting the rest of blend  while keeping the Mg I contribution
fixed. N$_{\rm Cr \ II}$ was determined by simultaneously fitting the Cr II $\lambda$ 2056 line and the blended Cr II + Zn II $\lambda$ 2062 line, where the contribution from
Zn II was estimated from the Zn II + Mg I $\lambda$ 2026 line. See also \citet{Kh04} for a discussion of the profile fitting scheme. In this paper we adopt the standard
notation for relative abundance:
$$[X/Y] = log (N_{\rm X}/N_{\rm H \ I}) - log (X/H)_{\sun},$$
Solar system abundances have been adopted from \citet{Lodd03}.

In addition to the Voigt profile fitting method, the package SPECP, also developed by D.E. Welty, was used to determine column densities via the apparent optical depth method
(AOD) \citep{Sav96}. We used SPECP to measure the equivalent widths of various transitions as well. We present the rest-frame equivalent widths ($W_{0}$) of various lines in
Table \ref{ewtable}. The 1$\sigma$ errors for the equivalent widths are also given and include the effect of both, the photon noise and the uncertainty in continuum placement.
In the case of the non-detection of a line, the limiting equivalent width was determined from the local signal to noise ratio (S/N), and a corresponding 3$\sigma$ column density
upper limit was determined, assuming a linear curve of growth. Cells with ``..." entries represent lines which could not be measured due to one or a combination of the following:
lack of coverage, blending with Ly$\alpha$ forest lines, blending with atmospheric absorption bands, very poor S/N due to spectrograph inefficiency at wavelength extremes and
coincidence of the line with damaged portions of the CCD.

\begin{table*}
\begin{minipage}{\textwidth}
\setlength{\tabcolsep}{5.3pt}
\setlength{\extrarowheight}{1.75pt}
\footnotesize
\caption{Rest-frame equivalent widths of key metal lines from this sample. Measured values and 1$\sigma$ errors are in m\ang units.}
\label{ewtable}
\begin{tabular*}{\textwidth}{ccccccccccccc}

\hline
\hline
QSO		&	z$_{abs}$	&	Mg I		&	Mg II		&	Mg II		&	Al II		&	Al III		&	Al III		&	S II		&	S II		&	S II		&	Si II		&	Si II		\\
		&			&	2852		&	2796		&	2803		&	1670		&	1854		&	1862		&	1250		&	1253		&	1259		&	1526		&	1808		\\
\hline																									
Q1039-2719	&	2.139		&	749$\pm$11	&	1527$\pm$54	&	1365$\pm$66	&	572$\pm$14	&	$<$933$^{b}$	&	212$\pm$8	&	39$\pm$4	&	$<$111$^{b}$	&	103$\pm$3	&	529$\pm$16	&	108$\pm$12	\\
Q1103-2645	&	1.839		&	86$\pm$10	&	1034$\pm$10	&	768$\pm$12	&	...		&	67$\pm$10	&	...		&	6$\pm$3		&	$<$42$^{b}$	&	12$\pm$4	&	213$\pm$5	&	$<$5		\\
Q1311-0120	&	1.762		&	213$\pm$73	&	2230$\pm$53	&	1601$\pm$53	&	...		&	$<$12		&	$<$12		&	...		&	...		&	...		&	$>$306$^{a}$	&	...		\\
Q1551+0908	&	2.320		&	...		&	...		&	...		&	114$\pm$3	&	23$\pm$7	&	...		&	21$\pm$3	&	...		&	...		&	120$\pm$10	&	5$\pm$2		\\
\vspace{0.75mm}
Q2123-0050	&	2.058		&	527$\pm$16	&	1946$\pm$6	&	1655$\pm$6	&	654$\pm$10	&	197$\pm$12	&	103$\pm$13	&	70$\pm$6	&	196$\pm$10	&	...		&	546$\pm$9	&	40$\pm$6	\\

\hline																									
QSO		&	z$_{abs}$	&	Cr II		&	Mn II		&	Mn II		&	Mn II		&	Fe II		&	Fe II		&	Fe II		&	Fe II		&	Fe II		&	Zn II$^{c}$	&	Zn II$^{d}$	\\
		&			&	2056		&	2576		&	2594		&	2606		&	2344		&	2374		&	2382		&	2586		&	2600		&	2026		&	2062		\\
\hline																									
Q1039-2719	&	2.139		&	36$\pm$6	&	70$\pm$9	&	...		&	$<$113$^{b}$	&	568$\pm$5	&	414$\pm$15	&	747$\pm$15	&	575$\pm$13	&	1478$\pm$244	&	48$\pm$12	&	23$\pm$4	\\
Q1103-2645	&	1.839		&	$<$3		&	$<$46$^{b}$	&	$<$21$^{b}$	&	$<$6		&	139$\pm$6	&	56$\pm$12	&	327$\pm$9	&	124$\pm$10	&	303$\pm$7	&	$<$4		&	$<$3		\\
Q1311-0120	&	1.762		&	25$\pm$5	&	$<$10		&	$<$93$^{b}$	&	...		&	384$\pm$49	&	154$\pm$25	&	766$\pm$53	&	305$\pm$53	&	$<$1281$^{b}$	&	$>$94		&	$>$26		\\
Q1551+0908	&	2.320		&	$<$6		&	$<$5		&	...		&	...		&	129$\pm$5	&	44$\pm$4	&	250$\pm$10	&	193$\pm$9	&	273$\pm$7	&	$<$4		&	$<$4		\\
\vspace{0.75mm}
Q2123-0050	&	2.058		&	$<$3		&	23$\pm$3	&	$<$4		&	15$\pm$6	&	445$\pm$8	&	221$\pm$72	&	862$\pm$216	&	328$\pm$20	&	772$\pm$16	&	47$\pm$6	&	21$\pm$18	\\
\hline

\end{tabular*}
$^{a}$This line is partially blended with Ly$\alpha$ forest lines. $^{b}$This line is blended with another feature. $^{c}$This line is blended with Mg I $\lambda$ 2026. Therefore, the measured value represents the total equivalent width of the blend. However, the Mg I contribution is judged to be insignificant in all cases.
$^{d}$Since this line is blended with the Cr II $\lambda$ 2062 line, the measured value represents the total equivalent width of the blended feature.\\
\end{minipage}
\end{table*}

\section{Discussion of Individual Objects}

\subsection{Q1039-2719, $z_{em}$=2.193}

\begin{figure}

\begin{center}

\includegraphics[angle=90, width=\linewidth]{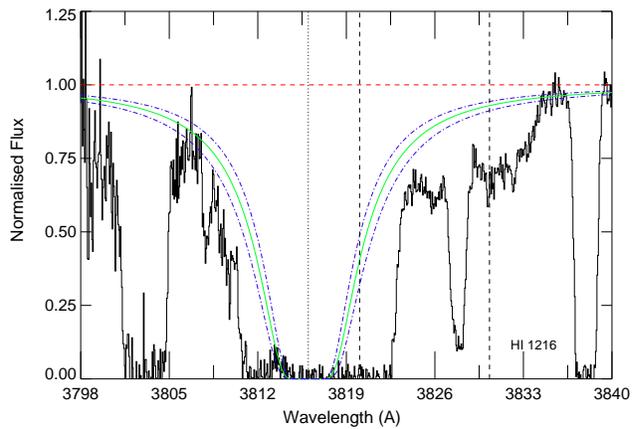}

\end{center}

\caption[Fig1]{Lyman-$\alpha$ absorption feature in the $z_{abs}$ = 2.139 system towards Q1039-2719. The solid green curve is the Voigt
               profile for log N$_{\rm H \ I}$ = 19.55. The blue dotted and dashed curves above and below the green curve are Voigt
	       profiles for log N$_{\rm H \ I}$ = 19.40 and 19.70, respectively. The red dashed line represents the normalized continuum
	       while the black dotted line denotes the profile center. The vertical dashed lines denote the locations of the N V
	       $\lambda\lambda$ 1239, 1243 lines from the $z_{abs}$ = 2.082 absorber. \label{q1039lyalphafig}}

\end{figure}

The sightline to this moderately bright BAL QSO traces a strong sub-DLA system at $z_{abs}$ = 2.139 in addition to a weak absorber at $z_{abs}$ = 2.082 and
three broad absorption systems at $z_{abs}$ = 1.518, 1.702, 1.757 \citep{SP01}. The continuum around the
Lyman-$\alpha$ line of the sub-DLA is affected by Si IV absorption from the BAL systems at $z_{abs}$ = 1.702 and 1.757 as well as N V
absorption from the $z_{abs}$ = 2.082 absorber. A relatively un-affected part of the spectrum redward of the Lyman-$\alpha$ line was
used to constrain the continuum. We made use of the residual flux at $\sim $3815 \ang to eliminate contribution from the  N V $\lambda\lambda$ 1239,
1243 lines in the $z_{abs}$ = 2.082 absorber as well as from the Ly$\alpha$ forest and to estimate log N$_{\rm H \ I}$ = 19.55$\pm$0.15.
The Voigt profile fit to the Lyman-$\alpha$ line is shown in Figure \ref{q1039lyalphafig}.

\begin{figure*}

\begin{center}

\includegraphics[angle=90, width=\linewidth]{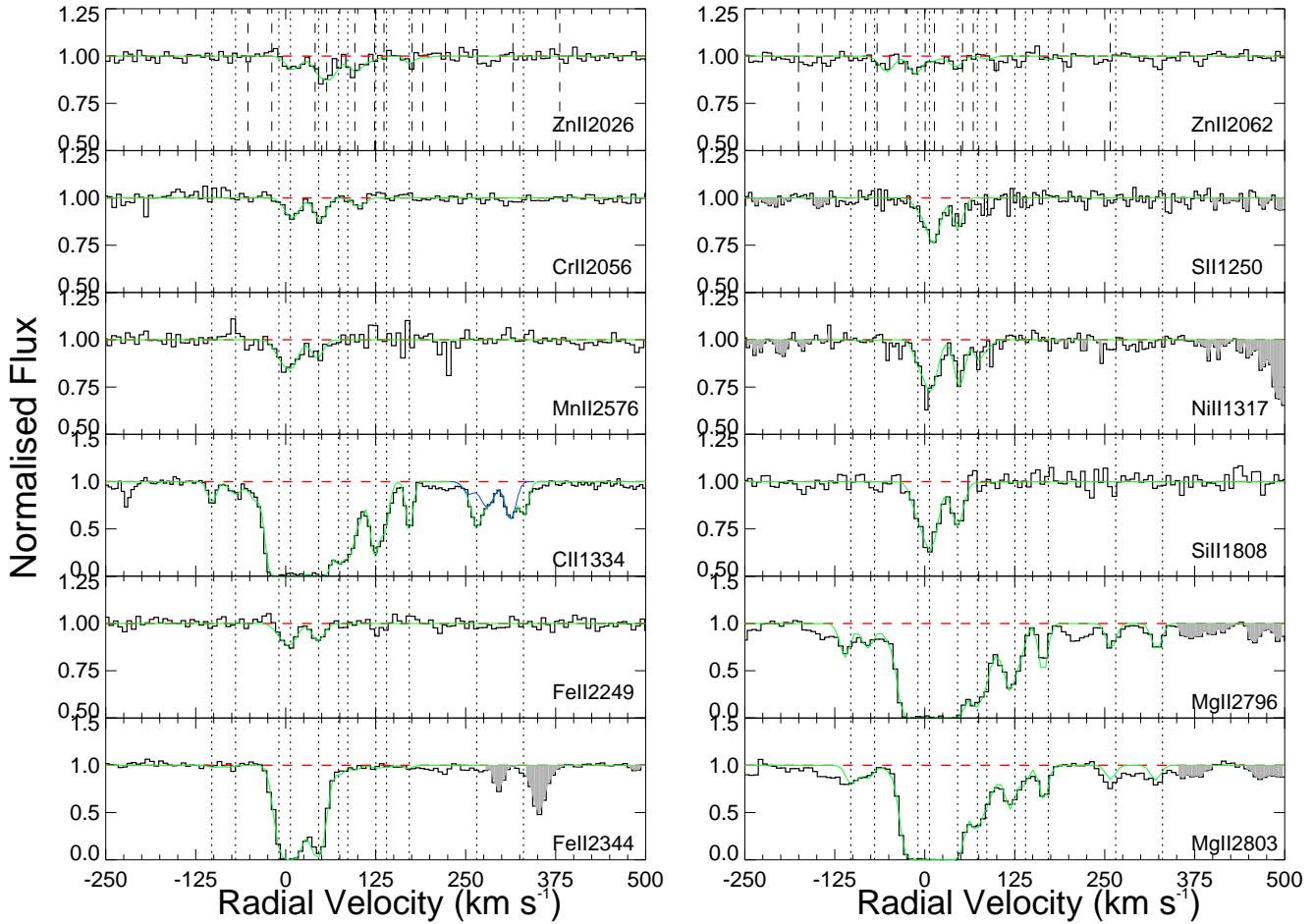}

\end{center}

\caption[Fig2]{Velocity plots for several lines of interest in the z =2.139 system in the spectrum of Q1039-2719. The solid green line indicates
the theoretical profile fit to the spectrum, and the dashed red line is the continuum level. The vertical dotted lines indicate the positions of
the components that were used in the fit. In the cases of the Zn II $\lambda\lambda$ 2026,2062 lines, which have other lines nearby, the long dashed
vertical lines indicate the positions of the components for Mg I (former case), and Cr II (latter case). The regions shaded in gray in some of the
panels represent features unrelated to the absorption systems presented here. In the ``CII 1334" panel, the solid green line represents the blend
between C II $\lambda$ 1334.5 and  C II$^{*} \, \lambda$ 1335.7 lines while the solid blue line represents the contribution from C II$^{*} \,
\lambda$ 1335.7 to this blend.\label{q1039ionsfig}}

\end{figure*}

The absorption profiles of this sub-DLA system show three strong components at velocities -9, 10, and 46 \kms along with several weak satellites spanning
a total $\sim$ 430 \kms. The sub-DLA is detected in absorption form several elements in multiple ionization stages such Mg I, Mg II, Fe II, Fe III, Si II,
Si III, Si IV, C I, C II, C IV, Al II, Al III, P II, Cr II, Mn II, Ni II, S II and Zn II. Table \ref{q1039table} shows the column densities in individual
velocity components for various ions. The Voigt profile fits to some of the lines of interest are shown in Figure \ref{q1039ionsfig}. It is to be noted
that, abundance measurements for various elements in this absorber have previously been reported by \citet{SP01}. However, their results included measurements
from the two strongest absorption components only and contributions from the weaker components, although small, were ignored. Therefore, the abundances were
affected by underestimation of column densities of various ions, including Zn~II and S II. To check the consistency of our abundance determinations from the
MIKE spectra, we derived column densities of various ions (e.g., log N$_{\rm S II}$ = 14.76$\pm$0.09, log N$_{\rm Fe II}$ = 14.69$\pm$0.06, log N$_{\rm Si II}$
= 14.99$\pm$0.01) using AOD measurements on the UVES spectra from \citet{SP01} and compared them with our results. For most of the ions, the column densities
agree within 1$\sigma$ uncertainties. We also detect C II* $\lambda$ 1335.7 in this sub-DLA, but the components of C II* at velocities -9 and 10 \kms
are blended with the C II $\lambda$ 1334 line in our MIKE spectrum. Although, we were able to measure the contribution from the component at 10 \kms using the
higher resolution UVES data from \citet{SP01}, the component at -9 \kms could not be separated from the blend, resulting in the placement of only a lower
limit on the abundance of C~II*. The C II* column densities listed in table \ref{q1039table} are from our measuremnts on the UVES data.

\begin{table*}
\begin{center}
\footnotesize
\caption{Column densities in individual velocity components for the $z$=2.139 absorber with log N$_{\rm H \ I}$=19.55 in Q1039-2719. Velocities and b$_{eff}$ values are given
         in units of \kms. Column densities are in units of cm$^{-2}$ and 1$\sigma$ errors in column densities are given.} 
\label{q1039table}
\begin{tabular}{cccccccc}
\hline\hline
Vel	&b$_{eff}$	&		Mg I		&		Mg II		&		Fe II		&		Zn II		&		Ni II		&		C II*		\\
\hline															
-103	&	6.6	&		-		&	(1.79$\pm$0.57)E+12	&	(5.28$\pm$3.04)E+11	&		-		&		-		&		-		\\
-70	&	9.5	&		-		&	(1.42$\pm$0.50)E+12	&	(6.10$\pm$3.21)E+11	&		-		&		-		&		-		\\
-42	&	10.4	&	(7.23$\pm$2.83)E+11	&	(1.05$\pm$0.65)E+12	&		-		&		-		&		-		&		-		\\
-9	&	8.7	&	(9.03$\pm$4.13)E+11	&	$>$6.59E+14		&	(7.37$\pm$0.51)E+13	&		-		&	(7.22$\pm$2.36)E+12	&	-$^{a}$ 		\\
10	&	11.5	&	(4.90$\pm$1.69)E+12	&	(2.02$\pm$0.64)E+15	&	(2.59$\pm$0.35)E+14	&	(5.31$\pm$1.32)E+11	&	(2.61$\pm$0.38)E+13	&	(1.89$\pm$0.27)E+13	\\
46	&	8.5	&	(2.47$\pm$0.87)E+12	&	$>$4.06E+14		&	(1.85$\pm$0.31)E+14	&	(4.42$\pm$1.26)E+11	&	(1.89$\pm$0.33)E+13	&	(2.02$\pm$0.26)E+13	\\
73	&	6.2	&	(3.28$\pm$2.56)E+11	&	$>$8.77E+12		&	(2.46$\pm$0.25)E+12	&		-		&	(5.63$\pm$1.92)E+12	&		-		\\
86	&	8.0	&		-		&	$>$4.60E+12		&	(1.38$\pm$0.25)E+12	&	(2.23$\pm$1.17)E+11	&	(3.00$\pm$1.96)E+12	&		-		\\
104	&	9.0	&	(3.86$\pm$2.77)E+11	&	(1.78$\pm$0.61)E+12	&		-		&		-		&		-		&		-		\\
125	&	4.4	&		-		&	(4.71$\pm$1.09)E+12	&	(6.12$\pm$1.66E+11	&		-		&		-		&		-		\\
140	&	6.4	&	(3.06$\pm$2.04)E+11	&	(1.91$\pm$0.59)E+12	&	(5.06$\pm$1.68)E+11	&		-		&		-		&		-		\\
172	&	3.6	&		-		&	(3.80$\pm$0.99)E+12	&	(7.94$\pm$1.55)E+11	&	(2.46$\pm$1.04)E+11	&		-		&		-		\\
265	&	5.6	&		-		&	(1.19$\pm$0.49)E+12	&	(4.40$\pm$1.41)E+11	&		-		&		-		&		-		\\
330	&	6.1	&		-		&	(1.31$\pm$0.50)E=12	&	(3.4$\pm$1.41)E+11	&		-		&		-		&		-		\\

\hline
\end{tabular}
\end{center}
$^{a}$This component is blended with the C II $\lambda$ 1334.5 line.\\ 

\end{table*}

Photoionisation calculations for this system, as described in $\S$ 5.2, suggest that the observed metallicity ([Zn/H] = $-$0.02 dex) and depletion ([Zn/Fe] = $+$0.28 dex)
underestimate the true values significantly. The corrected values for [Zn/H] and [Zn/Fe] were estimated to be +0.46 dex and +0.95 dex, respectively.

\subsection{Q1103-2645, $z_{em}$ = 2.145}

\begin{figure}

\begin{center}

\includegraphics[angle=90, width=\linewidth]{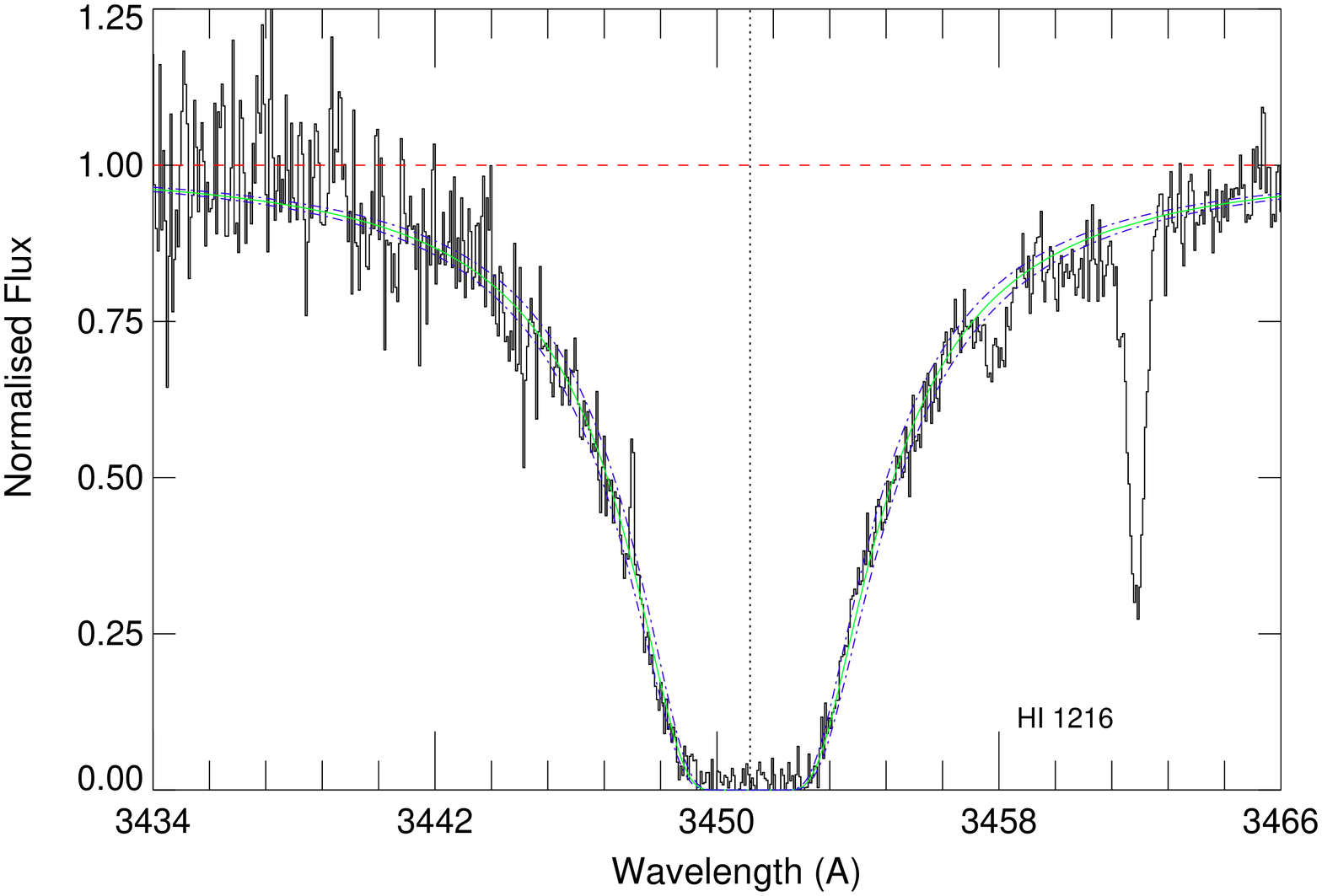}

\end{center}

\caption[Fig3]{Same as Fig. 1, but for the Lyman-$\alpha$ line in the $z_{abs}$ = 1.839 system towards Q1103-2654. The solid green
	       curve shows the Voigt profile for log N$_{\rm H \ I}$ = 19.52 while the blue dotted-dashed curves represent an
	       uncertainty of 0.04. \label{q1103lyalphafig}}

\end{figure}

This QSO sightline probes a sub-DLA at $z$ = 1.839 \citep{Petit00}. We estimate log N$_{\rm H \ I}$ = 19.52$\pm$0.04 for the absorber
by fitting a Voigt profile to the Lyman-$\alpha$ line (see Figure \ref{q1103lyalphafig}). Absorption features of various elements in different
ionisation stages such as Mg I, Mg II, Fe II, C II, C II*, S II, Si II, Si IV and Mn II, were detected in this system. The absorption profiles
reveal 11 components ranging from -163 \kms to 39 \kms but most of the absorption comes from two main components at -49 and -12 \kms. Several
key lines such as C IV $\lambda\lambda$ 1548, 1550; Al II$\lambda$ 1670 and Ni II $\lambda$ 1741 fell on a damaged portion near the red end of
the blue CCD of MIKE, preventing us from making reliable determination of column densities.

\begin{figure*}

\begin{center}

\includegraphics[angle=90, width=\linewidth]{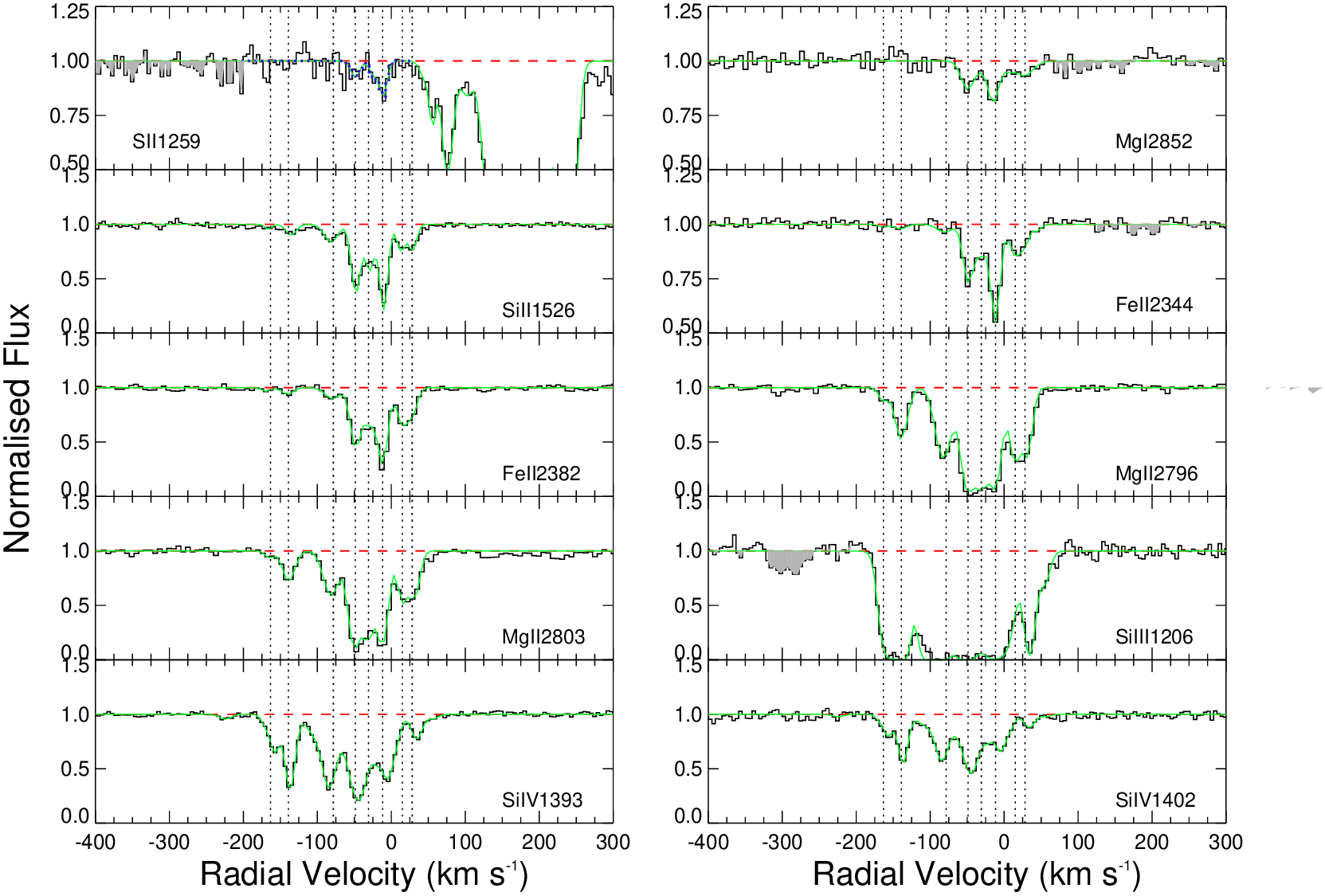}

\end{center}

\caption[Fig4]{Same as Fig. 2, but for the z$_{abs}$=1.839 system in the spectrum of Q1103-2654. In the ``SII 1259" panel, the solid green line
represents the total contribution from the S II $\lambda$ 1259.5 and the Si II $\lambda$ 1260.4 lines. The contribution from S II $\lambda$ 1259.5
alone is shown with the blue dotted line.\label{q1103ionsfig}}

\end{figure*}

Table \ref{q1103table} summarizes the results from profile fitting analysis for this system and the velocity plots for some of the lines of interest
are shown in Figure \ref{q1103ionsfig}. There was no detection of Zn with a S/N $\sim$ 45 near Zn II $\lambda$ 2026. Based on the 3$\sigma$ limiting
rest equivalent width, $W_{rest}$ = 3.9 m$\ang$, we estimate log N$_{\rm Zn \ II}$ $<$ 11.3 and [Zn/H] $<$ -0.82 for this absorber. S II was detected
in this system with log N$_{\rm S \ II}$ = 13.9 and [S/H] = -0.82. We note that, [S/H] for this system has also been reported by \citealt{Petit00}
and their value of -0.94$\pm$0.16 is consistent with our measurement within 1$\sigma$ uncertainties. Ionisation modelling for this absorber indicates
a moderate correction of -0.3 dex in S abundance (see section 5.2 for details). The data also show presence of Mn II $\lambda$ 2576 but this line is
blended with an unidentified feature. Since no other Mn II lines were detected, we could only place an upper limit on Mn abundance of this absorber.

\begin{table*}
\center
\footnotesize
\caption{Same as Table 3, but for the z$_{abs}$=1.839 absorber with log N$_{\rm H \ I}$=19.52 in Q1103-2645}
\label{q1103table}
\begin{tabular}{ccccccc}

\hline\hline
Vel	&b$_{eff}$	&		Mg I		&		Mg II		&		Fe II		&		C II		&		S II		\\
\hline															
-163	&	2.7	&		-		&	(4.54$\pm$6.74)E+11	&	(2.48$\pm$7.14)E+11	&	(3.46$\pm$2.05)E+12	&		-		\\
-139	&	8.2	&		-		&	(2.92$\pm$0.22)E+12	&	(6.08$\pm$8.34)E+11	&	(2.12$\pm$0.33)E+13	&		-		\\
-78	&	11.7	&		-		&	(5.86$\pm$0.28)E+12	&	(1.31$\pm$0.16)E+12	&	(4.17$\pm$0.34)E+13	&		-		\\
-67	&	2.2	&		-		&		-		&		-		&	(1.19$\pm$0.21)E+13	&		-		\\
-49	&	5.9	&	(2.07$\pm$0.33)E+11	&	$>$3.09E+13		&	(7.77$\pm$0.48)E+12	&	$>$3.44E+14		&	(2.02$\pm$1.72)E+13	\\
-31	&	6.1	&	(4.46$\pm$3.07)E+10	&	$>$1.60E+13		&	(3.52$\pm$0.40)E+12	&	$>$1.89E+14		&	(1.82$\pm$1.78)E+13	\\
-12	&	4.5	&	(2.79$\pm$0.34)E+11	&	$>$5.45E+13		&	(1.56$\pm$0.08)E+13	&	$>$3.34E+14		&	(3.97$\pm$1.88)E+13	\\
4	&	5.6	&	(6.67$\pm$3.01)E+10	&		-		&		-		&	(1.73$\pm$0.11)E+13	&		-		\\
15	&	5.8	&		-		&	$>$5.27E+12		&	(3.60$\pm$0.21)E+12	&	$>$2.68E+13		&		-		\\
28	&	4.7	&	(1.12$\pm$0.37)E+11	&	(4.33$\pm$0.33)E+12	&	(2.22$\pm$0.18)E+12	&	(2.19$\pm$0.13)E+13	&		-		\\
39	&	6.0	&		-		&		-		&		-		&	(1.36$\pm$0.09)E+13	&		-		\\
\hline

\end{tabular}
\end{table*}

\subsection{Q1311-0120, $z_{em}$=2.584}

This QSO sightline has a sub-DLA absorber, identified in the LBQS survey \citep{Wol95}, at $z$ = 1.762 with Lyman-$\alpha$ rest-frame equivalent width of
7.3$\pm$0.7 \ang. The Lyman-$\alpha$ line was partially covered in the extreme blue order of our echelle data and because of the very poor S/N in that
wavelength region, neutral hydrogen column density could not be determined using a Voigt profile fit. Instead, we estimate log N$_{\rm H \ I}$ =
20.00$\pm$0.08 from the rest-frame equivalent width reported by \citet{Wol95}, using the curve of growth for the H I Lyman-$\alpha$ line. This absorber
shows a relatively complex velocity structure and 12 components, spanning $\sim$ 550 \kms in velocity space were required to fit the observed absorption
profiles. While most of the absorption occurs in two component clusters appearing between -5 \kms and 200 \kms, a weaker absorption complex, separated from
the main components by more than 500 \kms, is detected in most of the strong transitions. Additional weaker components, bridging the gap between the satellite
and the main absorption, are seen only in the strongest of transitions such as Fe II $\lambda$ 2382 and Mg II $\lambda\lambda$ 2796, 2803. Results from the
profile fitting analysis for this system are shown in Table \ref{q1311table}.

\begin{table*}
\begin{center}
\footnotesize
\caption{Same as Table 3, but for the z$_{abs}$=1.762 absorber with N$_{\rm H \ I}$=20.00 in Q1311-0120}
\label{q1311table}
\begin{tabular}{ccccccc}
\hline\hline
Vel	&b$_{eff}$	&		Mg I		&		Mg II		&		Fe II		&		Zn II		&		Cr II		\\
\hline													
-5	&	3.8	&	(1.80$\pm$0.63)E+11	&	$>$5.93E+012		&	(6.07$\pm$1.40)E+12	&		-		&		-		\\
24	&	3.8	&	(5.01$\pm$5.24)E+10	&	$>$4.71E+013		&	(2.40$\pm$0.32)E+13	&		-		&		-		\\
45	&	6.1	&	(6.25$\pm$0.86)E+11	&	$>$1.80E+014		&	(1.04$\pm$0.13)E+14	&	(1.22$\pm$0.91)E+012	&	(8.72$\pm$2.44)E+12	\\
115	&	8.6	&	(1.16$\pm$0.63)E+10	&	(3.06$\pm$0.68)E+12	&	(1.16$\pm$0.29)E+12	&	(6.28$\pm$0.86)E+011	&		-		\\
169	&	8.0	&	(8.87$\pm$6.03)E+10	&	$>$2.97E+013		&	(1.44$\pm$0.15)E+13	&		-$^{a}$ 	&		-		\\
184	&	7.3	&	(3.07$\pm$0.70)E+11	&	$>$2.72E+013		&	(4.92$\pm$1.09)E+12	&		-$^{a}$ 	&		-		\\
244	&	3.9	&	(7.91$\pm$5.26)E+10	&	$>$1.55E+013		&	(3.09$\pm$0.41)E+12	&		-$^{a}$ 	&		-		\\
275	&	7.9	&		-		&	$>$9.13E+012		&	(1.89$\pm$0.32)E+12	&		-		&		-		\\
294	&	8.4	&		-		&	$>$8.57E+012		&	(1.33$\pm$0.32)E+12	&	(7.86$\pm$0.93)E+011	&		-		\\
317	&	8.5	&		-		&	(2.23$\pm$0.62)E+12	&	(1.30$\pm$0.31)E+12	&		-		&		-		\\
525	&	6.0	&	(2.06$\pm$0.61)E+11	&	$>$1.32E+013		&	(6.01$\pm$1.30)E+12	&		-		&		-		\\
546	&	2.9	&	(2.28$\pm$0.70)E+11	&	$>$4.40E+012		&	(2.01$\pm$0.37)E+12	&	(5.76$\pm$0.82)E+011	&		-		\\

\hline
\end{tabular}
\end{center}
$^{a}$This component is blended with an unidentified feature.\\
\end{table*}

\begin{figure*}

\begin{center}

\includegraphics[angle=90, width=\linewidth]{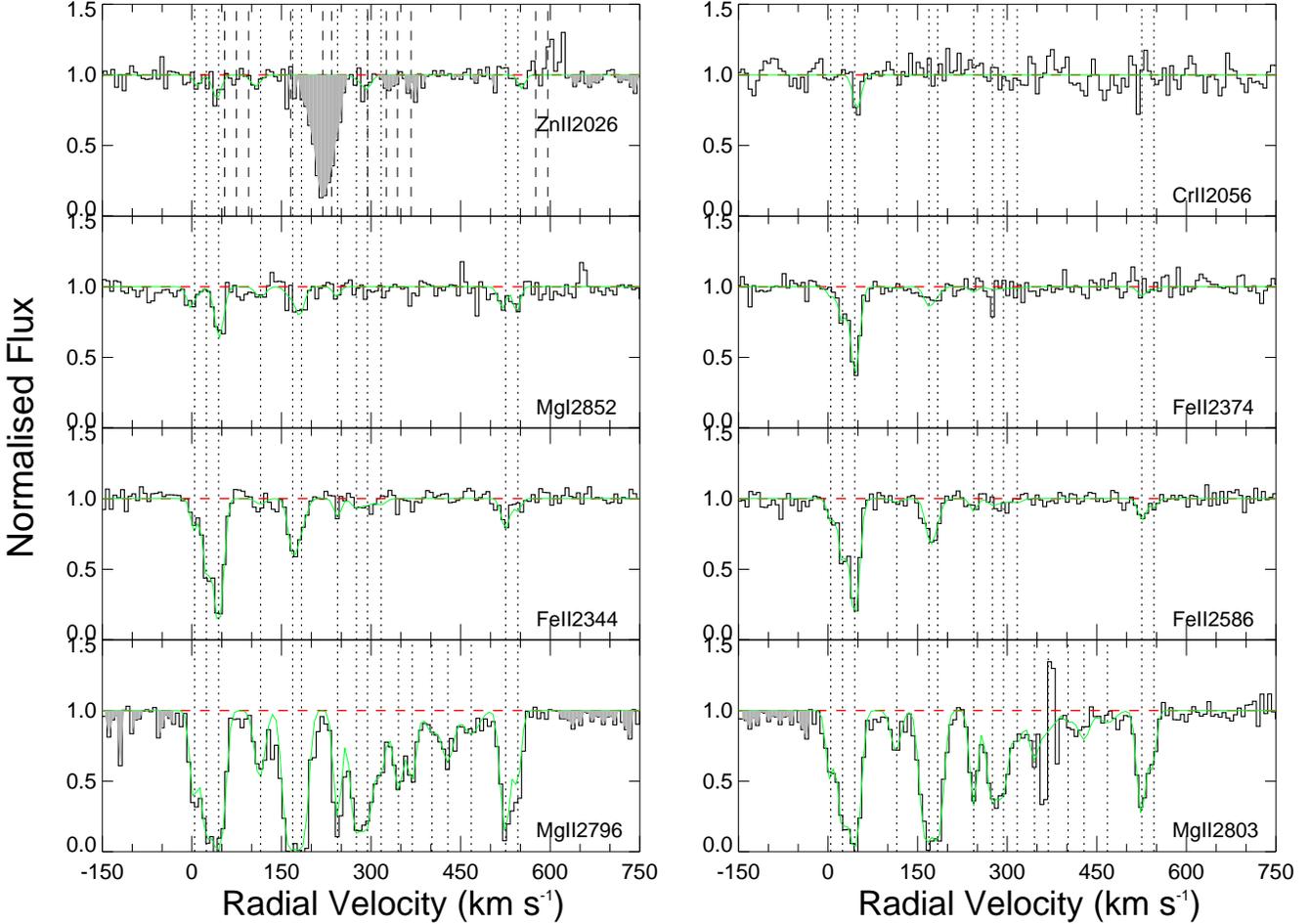}

\end{center}

\caption[Fig5]{Same as Fig. 2, but for the z$_{abs}$=1.762 system in the spectrum of Q1311-0120. \label{q1311ionsfig}}

\end{figure*}

Our data near the extreme blue end of the spectral coverage were affected by poor S/N owing to the combination of lower sensitivity and continuum absorption
form the Ly$\alpha$ forest clouds. Several of the lines of interest such as S II $\lambda\lambda\lambda$ 1250, 1253, 1259; Si II $\lambda$ 1304; Ni II
$\lambda\lambda$ 1317, 1370; C II $\lambda$ 1334; C II* $\lambda$ 1336 and Si IV $\lambda\lambda$ 1393, 1402 were located in this region and therefore could
not be analysed reliably. Due to the high redshift of the background QSO, even lines with higher rest wavelengths such as C IV $\lambda\lambda$ 1548, 1550
were blended with Ly$\alpha$ forest lines. Si II $\lambda$ 1526 was partly blended with forest lines resulting in the placement of only a lower limit on Si
II abundance. Zn II $\lambda$ 2026 line was detected in several components in this system. However, a part of the core component structure of the line
is blended with a strong unidentified feature and therefore, we report only a lower limit of log N$_{\rm Zn \ II}$ $>$ 12.57 and [Zn/H] $>$ -0.06, based on
the measurements of the un-blended components. The component at 546 \kms, unlikely to be associated with the main absorber galaxy, contributes $\sim$ 15$\%$
of the observed Zn II column density. However, the system is found to be metal rich ([Zn/H] $>$ -0.14) even if contribution from this high-velocity component
is ignored. This near-solar metallicity absorber also shows a high depletion with [Zn/Fe] $>$ +1.18. Ni II $\lambda$ 1741 and Al II $\lambda$ 1670 were
affected by cosmetic defect in the chip, however, we were able to place a lower limit on Al II abundance based on unaffected regions in the line. Velocity plots
for several lines of interest are shown in Figure \ref{q1311ionsfig}.

\subsection{Q1551+0908, $z_{em}$=2.739}

\begin{figure}

\begin{center}

\includegraphics[angle=90, width=\linewidth]{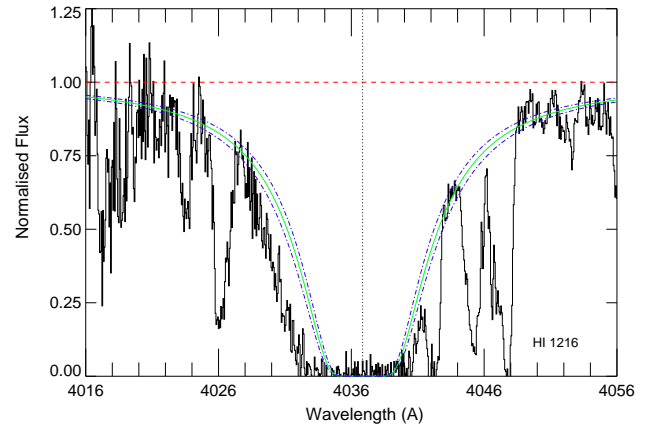}

\end{center}

\caption[Fig6]{Same as Fig. 1, but for the Lyman-$\alpha$ line in the $z_{abs}$ = 2.320 system towards Q1551+0908. The solid green curve shows the
	       Voigt profile for log N$_{\rm H \ I}$ = 19.70 while the blue dotted-dashed curves represent an uncertainty of 0.05. \label{q1551lyalphafig}}

\end{figure}

\begin{figure*}

\begin{center}

\includegraphics[angle=90, width=\linewidth]{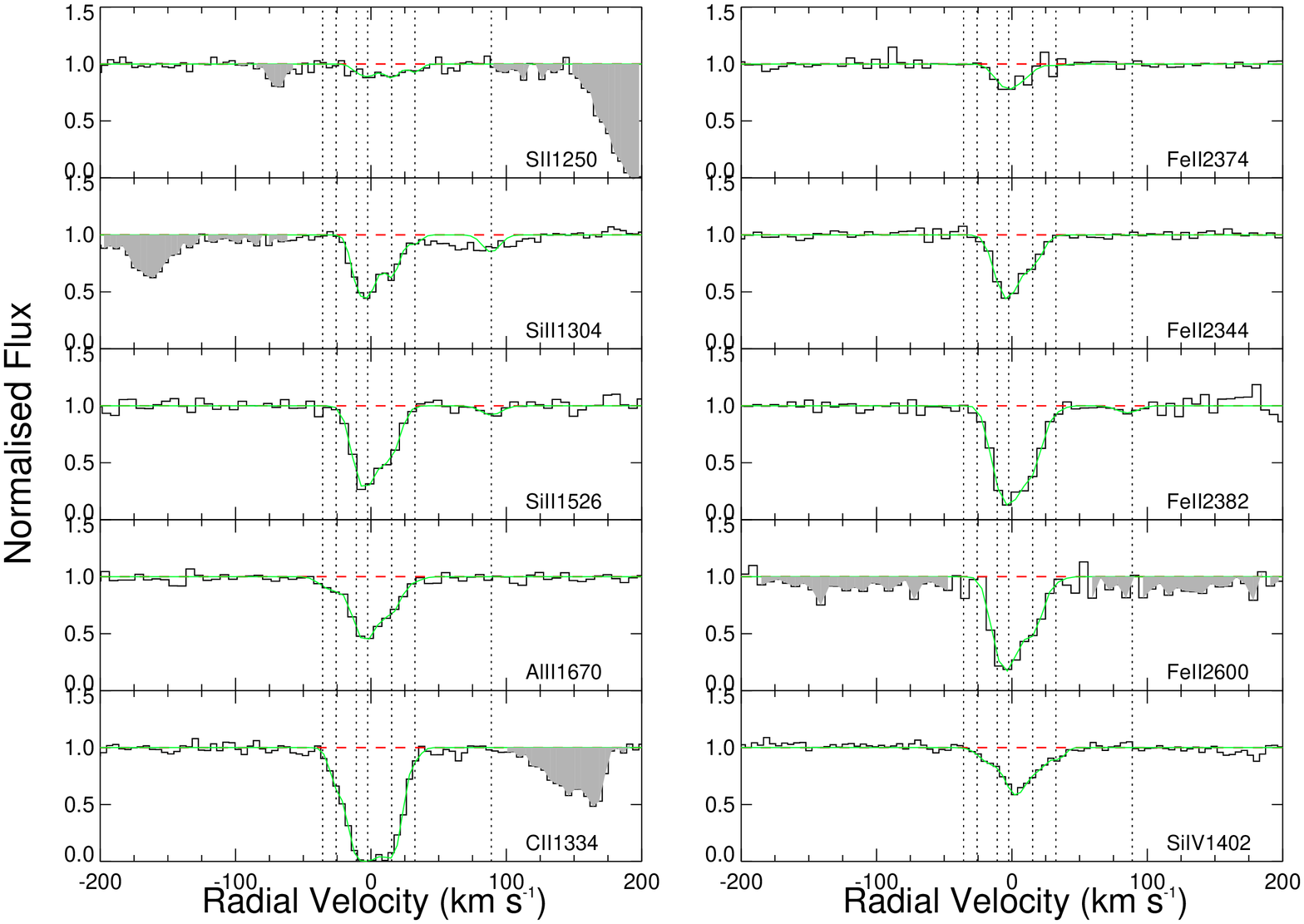}

\end{center}

\caption[Fig7]{Same as Fig. 2, but for the z$_{abs}$= 2.320 system in the spectrum of Q1551+0908. \label{q1551ionsfig}}

\end{figure*}

This QSO sightline has a sub-DLA absorber at $z$ = 2.320 \citep{Not09}. A Voigt profile fit to the Lyman-$\alpha$ line, shown in Figure
\ref{q1551lyalphafig}, yields log N$_{\rm H \ I}$=19.70$\pm$0.05. This sub-DLA is detected in absorption from Fe II, Fe III, Si II, Si IV,
C II, C III, C IV, Al II, Al III, S II, and Ni II. Mg I $\lambda$ 2852 and Mg II $\lambda\lambda$ 2796, 2803 were not covered. The observed
absorption profiles show a relatively simple velocity structure for this system requiring 5 components for an adequate fit. Table \ref{q1551table}
shows results from profile fitting analysis for this absorber. Zn II $\lambda$ 2026 was not detected in our data with S/N $\sim$ 40 near the
expected position of the line. Our estimate of a 3 $\sigma$ limiting rest-frame equivalent width of $W_{rest}$ = 4.38 m\ang places an upper
limit on the Zn II column density at log N$_{\rm Zn \ II}$ $<$ 11.38 and [Zn/H] $<$ -0.95. Measurement of the detected S II lines yield [S/H]
= -0.46 and suggest significant $\alpha$-enhancement with [S/Zn] $>$ 0.49. Figure \ref{q1551ionsfig} shows velocity plots for several lines of
interest along with their Voigt profile fits.

\begin{table*}
\center
\footnotesize
\caption{Same as Table 3, but for the z$_{abs}$=2.320 absorber with  log N$_{\rm H \ I}$=19.70 in Q1551+0908}
\label{q1551table}
\begin{tabular}{ccccccc}

\hline\hline
Vel	&b$_{eff}$	&		Fe II		&		Si II		&		C II		&		Al II		&		S II		\\
\hline													
-11	&	3.1	&	(4.54$\pm$0.94)E+012	&	(1.46$\pm$0.49)E+13	&	(3.49$\pm$2.20)E+14	&	(6.00$\pm$1.13)E+11	&		-		\\
-2	&	7.0	&	(1.99$\pm$0.28)E+013	&	(3.93$\pm$0.48)E+13	&		$>$2.37E+14	&	(1.73$\pm$0.13)E+12	&	(9.72$\pm$3.28)E+13	\\
15	&	5.8	&	(1.10$\pm$0.24)E+013	&	(2.20$\pm$0.31)E+13	&		$>$1.45E+14	&	(8.33$\pm$0.71)E+11	&	(1.02$\pm$0.28)E+14	\\
32	&	3.0	&	(3.21$\pm$2.26)E+011	&	(3.14$\pm$1.85)E+12	&	(2.49$\pm$1.04)E+12	&	(6.56$\pm$4.15)E+10	&	(4.72$\pm$2.38)E+13	\\
85	&	7.2	&	(6.04$\pm$2.57)E+011	&	(2.95$\pm$1.28)E+12	&		-		&		-		&		-		\\
\hline

\end{tabular}
\end{table*}

\subsection{Q2123-0050, $z_{em}$=2.262}

\begin{figure}

\begin{center}

\includegraphics[angle=90, width=\linewidth]{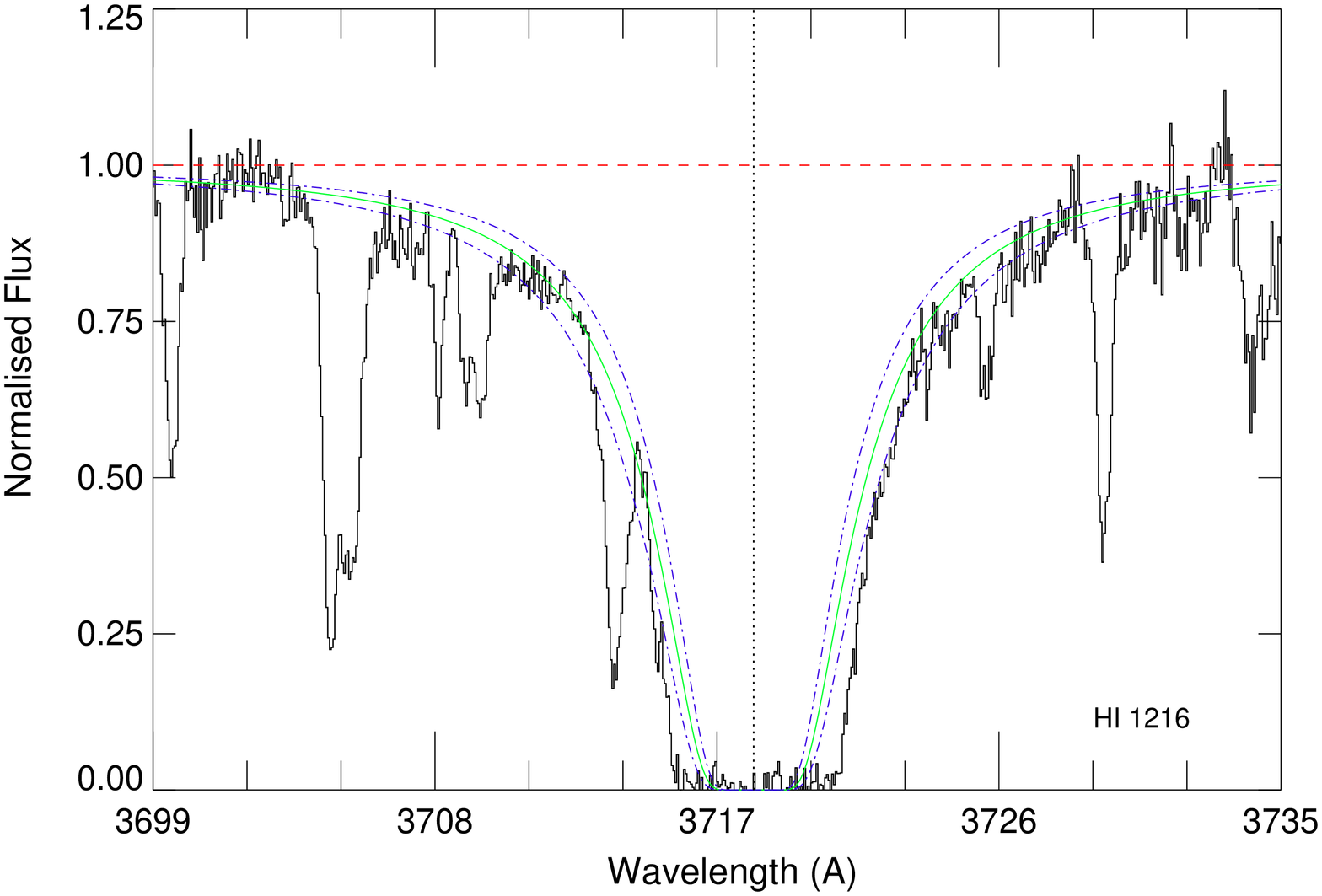}

\end{center}

\caption[Fig8]{Same as Fig. 1, but for the Lyman-$\alpha$ line in the $z_{abs}$ = 2.058 system towards Q2123-0050. The solid green curve shows the
	       Voigt profile for log N$_{\rm H \ I}$ = 19.35 while the blue dotted-dashed curves represent an uncertainty of 0.10. \label{q2123lyalphafig}}

\end{figure}

\begin{figure*}

\begin{center}

\includegraphics[angle=90, width=\linewidth]{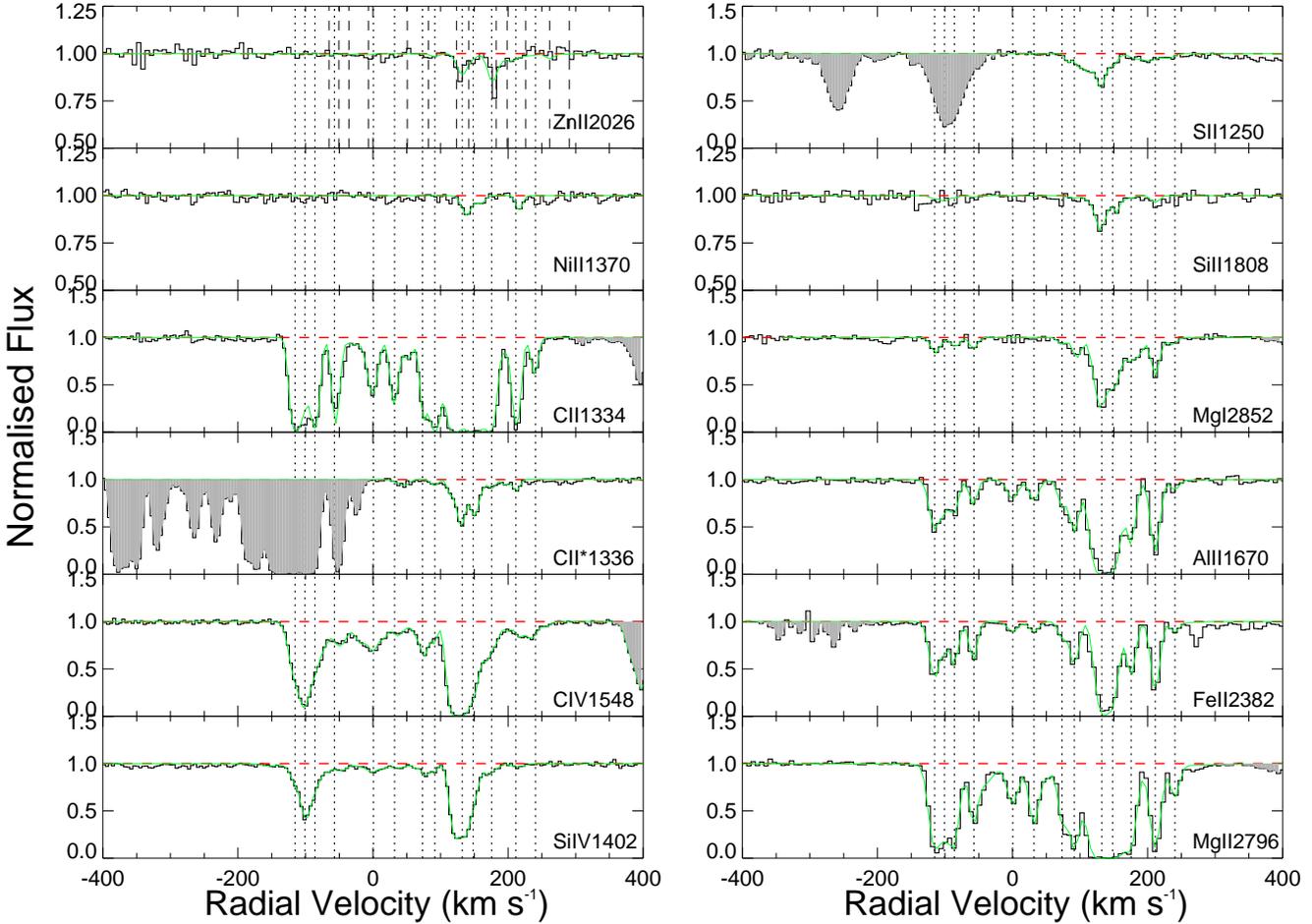}

\end{center}

\caption[Fig9]{Same as Fig. 2, but for the z$_{abs}$=2.058 system in the spectrum of Q2123-0050. In the ``CII$^{*}$1336" panel, the shaded region represents
absorption from the C II $\lambda$ 1334.5 line.\label{q2123ionsfig}}

\end{figure*}

This quasar sightline traces a sub-DLA at $z$ = 2.058 \citep{Kap10} with log N$_{\rm H \ I}$ = 19.35$\pm$0.10. Figure \ref{q2123lyalphafig} shows the
Voigt profile we used to determine the neutral hydrogen column density for this system. A complex structure with 13 components spanning more than 350
\kms in velocity was required to model the absorption characteristics of the sub-DLA. Details of the absorption structure analysis are given in Table
\ref{q2123table}. Absorption signatures from various ions such as Mg I, Mg II, Fe II, Si II, Si IV, Al II, Al III, C II, C II*, C IV, Mn II, Ni II, S
II and Zn II were detected in QSO spectrum at the sub-DLA redshift. Figure \ref{q2123ionsfig} shows the velocity plots of several lines of interest
along with their Voigt profile fits. The metallicity of this system, based on the observed Zn II column density of log N$_{\rm Zn \ II}$ = 12.23,
is super-solar ([Zn/H] = +0.25), making it the most metal-rich sub-DLA QSO absorber known so far at $z$ $>$ 2 (we note here, that higher metallicities
in some lower-redshift sub-DLAs have been reported by \citealt{Mei07,Mei08,Mei09,Per06a,Per08,Pro06}). In any case, due to the relatively low N$_{\rm H
\ I}$ of this absorber, it is necessary to explore the extent of ionisation effects on the metallicity value. Indeed, the observed high values of column
density ratios between adjacent ions such as Al III/Al II, Si III/Si II, and Al III/Fe II in this absorber suggest a high level of ionisation. Our
photoionisation calculations indicate a correction of +0.59 dex for [Zn/H] (See sec. 5.2 for further details).

\begin{table*}
\center
\footnotesize
\caption{Same as Table 3, but for the z$_{abs}$=2.058 absorber with log N$_{\rm H \ I}$=19.35 in Q2123-0050}
\label{q2123table}
\begin{tabular}{cccccccc}

\hline\hline
Vel	&b$_{eff}$	&		Mg I		&		Fe II		&		Si II		&		S II		&		Zn II		&		Mn II		\\
\hline															
-116	&	5.8	&	(2.20$\pm$0.29)E+11	&	(8.84$\pm$0.45)E+12	&	(2.69$\pm$1.10)E+13	&		-		&		-		&		-		\\
-101	&	1.4	&		-		&	(2.38$\pm$0.42)E+12	&	(4.04$\pm$8.30)E+13	&		-		&		-		&		-		\\
-86	&	5.1	&	(1.22$\pm$0.26)E+11	&	(5.16$\pm$0.38)E+12	&	(1.87$\pm$0.88)E+13	&		-		&		-		&		-		\\
-57	&	3.3	&	(1.71$\pm$0.43)E+11	&	(5.11$\pm$0.40)E+12	&	(1.07$\pm$0.73)E+13	&		-		&		-		&		-		\\
0	&	8.2	&		-		&	(1.08$\pm$0.26)E+12	&	(3.06$\pm$0.56)E+12	&		-		&		-		&		-		\\
32	&	5.6	&		-		&	(1.28$\pm$0.24)E+12	&	(4.87$\pm$0.56)E+12	&		-		&		-		&		-		\\
74	&	2.3	&	(8.35$\pm$2.6)E+10	&	(2.77$\pm$0.45)E+12	&	(6.76$\pm$0.73)E+12	&	(5.64$\pm$1.50)E+13	&		-		&		-		\\
91	&	5.1	&	(2.94$\pm$0.32)E+11	&	(7.65$\pm$0.53)E+12	&	(2.17$\pm$0.11)E+13	&	(1.19$\pm$0.17)E+14	&	(9.56$\pm$9.00)E+010	&		-		\\
128	&	6.6	&	(2.12$\pm$0.1E)+12	&	(3.35$\pm$0.30)E+13	&	(3.71$\pm$0.35)E+14	&	(3.87$\pm$0.22)E+14	&	(7.88$\pm$1.11)E+011	&	(4.17$\pm$1.40)E+011	\\
148	&	7.3	&	(1.11$\pm$0.06)E+12	&	(1.99$\pm$0.09)E+13	&	(8.51$\pm$4.10)E+13	&	(1.24$\pm$0.17)E+14	&		-		&	(5.16$\pm$1.44)E+011	\\
175	&	4.4	&	(3.21$\pm$0.35)E+11	&	(7.71$\pm$0.55)E+12	&	(2.06$\pm$0.12)E+13	&	(5.77$\pm$1.90)E+13	&	(6.48$\pm$1.08)E+011	&	(3.44$\pm$1.28)E+011	\\
212	&	4.1	&	(7.19$\pm$0.59)E+11	&	(2.69$\pm$0.17)E+13	&	(5.96$\pm$0.47)E+13	&	(1.12$\pm$0.10)E+14	&		-		&		-		\\
240	&	4.9	&	(9.43$\pm$2.6)E+10	&	(9.82$\pm$2.50)E+11	&	(2.36$\pm$0.49)E+12	&	(6.08$\pm$1.80)E+13	&		-		&		-		\\
\hline

\end{tabular}
\end{table*}

\section{Results}

\begin{table*}
\setlength{\tabcolsep}{4.7pt}
\setlength{\extrarowheight}{1.75pt}
\footnotesize
\caption{Total column densities for the absorbers in this sample. Cells with ``..." entries represent undetermined column densities.}
\label{totalNtable}
\begin{tabular*}{\textwidth}{ccccccccccc}

\hline\hline
QSO		&	z$_{abs}$	&	log N$_{\rm H \ I}$	&	log N$_{\rm Mg \ I}$	&	log N$_{\rm Mg  \ II}$	&	log N$_{\rm Al \ II}$	&	log N$_{\rm Al \ III}$	&	log N$_{\rm C \ II}$	&	log N$_{\rm C \ II*}$	&	log N$_{\rm C \ IV}$	&	log N$_{\rm S \ II}$	\\
		&			&		cm$^{-2}$	&		cm$^{-2}$	&		cm$^{-2}$	&		cm$^{-2}$	&		cm$^{-2}$	&		cm$^{-2}$	&		cm$^{-2}$	&		cm$^{-2}$	&		cm$^{-2}$	\\
\hline																					
Q1039-2719	&	2.139		&	19.55$\pm$0.15		&		13.02$\pm$0.09	&		$>$15.49	&		$>$13.88	&		13.53$\pm$0.02	&		$>$15.78	&		$>$13.35	&		...		&		14.76$\pm$0.03	\\
AOD		&			&				&		12.98$\pm$0.01	&		$>$15.16	&		$>$13.66	&		13.52$\pm$0.01	&		$>$15.08	&				&				&		14.75$\pm$0.04	\\
Q1103-2645	&	1.839		&	19.52$\pm$0.04		&		11.86$\pm$0.05	&		$>$14.08	&		...		&		12.74$\pm$0.07	&		$>$15.01	&		$>$12.93	&		...		&		13.89$\pm$0.17	\\
AOD		&			&				&		11.84$\pm$0.05	&		$>$13.79	&				&		12.64$\pm$0.06	&		$>$14.69	&		$>$12.30	&				&		13.73$\pm$0.14	\\
Q1311-0120	&	1.762		&	20.00$\pm$0.08		&		12.27$\pm$0.04	&		$>$14.55	&		$>$13.03	&		$<$11.85	&		...		&		...		&		...		&		...		\\
AOD		&			&				&		12.25$\pm$0.14	&		$>$14.15	&		$>$12.90	&				&				&				&				&				\\
Q1551+0908	&	2.320		&	19.70$\pm$0.05		&			...	&		...		&		12.55$\pm$0.02	&		12.05$\pm$0.08	&		$>$14.87	&		$<$12.17	&		13.81$\pm$0.02	&		14.43$\pm$0.09	\\
AOD		&			&				&				&				&		12.53$\pm$0.01	&		12.14$\pm$0.13	&		$>$14.51	&				&		13.77$\pm$0.03	&		14.47$\pm$0.07	\\
Q2123-0050	&	2.058		&	19.35$\pm$0.10		&		12.74$\pm$0.01	&		$>$14.50	&		$>$15.24	&		13.45$\pm$0.07	&		$>$15.99	&		$>$13.82	&		$>$14.62	&		15.05$\pm$0.02	\\
\vspace{0.75mm}
AOD		&			&				&		12.72$\pm$0.01	&		$>$14.27	&		$>$13.47	&		13.14$\pm$0.02	&		$>$15.16	&		$>$13.80	&		$>$14.57	&		15.01$\pm$0.04	\\

\hline																					
																					
QSO		&	z$_{abs}$	&	log N$_{\rm H \ I}$	&	log N$_{\rm Si \ II}$	&	log N$_{\rm Si \ III}$	&	log N$_{\rm Si \ IV}$	&	log N$_{\rm Cr \ II}$	&	log N$_{\rm Mn \ II}$	&	log N$_{\rm Ni \ II}$	&	log N$_{\rm Fe \ II}$	&	log N$_{\rm Zn \ II}$	\\
		&			&		cm$^{-2}$	&		cm$^{-2}$	&		cm$^{-2}$	&		cm$^{-2}$	&		cm$^{-2}$	&		cm$^{-2}$	&		cm$^{-2}$	&		cm$^{-2}$	&		cm$^{-2}$	\\
\hline																					
Q1039-2719	&	2.139		&	19.55$\pm$0.15		&	15.30$\pm$0.03		&		$>$14.48	&		$>$14.50	&		13.07$\pm$0.04	&		12.48$\pm$0.06	&		13.79$\pm$0.04	&		14.72$\pm$0.04	&		12.16$\pm$0.07	\\
AOD		&			&				&	15.31$\pm$0.05		&		$>$14.30	&		$>$14.50	&		12.99$\pm$0.07	&		12.54$\pm$0.06	&		13.75$\pm$0.07	&		14.74$\pm$0.02	&		12.40$\pm$0.08	\\
Q1103-2645	&	1.839		&	19.52$\pm$0.04		&	14.07$\pm$0.02		&		$>$14.64	&		13.84$\pm$0.01	&		$<$11.91	&		$<$12.46	&		$<$12.34	&		13.54$\pm$0.02	&		$<$11.33	\\
AOD		&			&				&	14.00$\pm$0.01		&		$>$14.12	&		13.82$\pm$0.01	&				&		$<$12.37	&				&		13.52$\pm$0.04	&				\\
Q1311-0120	&	1.762		&	20.00$\pm$0.08		&		$>$14.38	&		...		&		...		&		12.94$\pm$0.12	&		$<$11.65	&		...		&		14.23$\pm$0.35	&		$>$12.57	\\
AOD		&			&				&		$>$14.26	&				&				&		12.87$\pm$0.09	&				&				&		14.09$\pm$0.06	&		$>$12.75	\\
Q1551+0908	&	2.320		&	19.70$\pm$0.05		&	13.91$\pm$0.04		&				&		13.34$\pm$0.01	&		$<$12.15	&		$<$11.40	&		13.01$\pm$0.32	&		13.56$\pm$0.05	&		$<$11.38	\\
AOD		&			&				&	13.97$\pm$0.02		&		$>$13.69	&		13.31$\pm$0.02	&				&				&		13.11$\pm$0.04	&		13.57$\pm$0.07	&				\\
Q2123-0050	&	2.058		&	19.35$\pm$0.10		&	14.89$\pm$0.06		&		$>$14.72	&		$>$13.99	&		$<$11.90	&		12.11$\pm$0.08	&		13.12$\pm$0.08	&		14.09$\pm$0.01	&		12.23$\pm$0.06	\\
\vspace{0.75mm}
AOD		&			&				&	14.85$\pm$0.06		&		$>$14.15	&		$>$13.95	&				&		12.04$\pm$0.06	&		13.01$\pm$0.10	&		14.01$\pm$0.01	&		12.44$\pm$0.06	\\

\hline

\end{tabular*}
\end{table*}

\subsection{Total Column Densities}
   
The results of the Voigt profile fits to various absorption features from the sub-DLAs in this sample are summarised in Table \ref{totalNtable}. Log of the total
column densities (sum of the column densities in the individual components determined via the profile-fitting method) for various ions are listed in this table.
Column densities, determined using the apparent optical depth (AOD) method to check the consistency of our fits and are also listed in Table \ref{totalNtable}.
In most cases, the column densities from the profile fitting and AOD methods agree to within the error bars, especially for the weak and unsaturated lines. Cells
with ``..." entries have undetermined column densities due to the reasons described in $\S$ 3. Total Zn II column density from Table \ref{totalNtable} and the
corresponding N$_{\rm H \ I}$ value for an absorber were used to determine its metallicity, [Zn/H]. Abundances of S and Fe relative to H were determined likewise.
The metallicities and other relative abundances for the observed systems are listed in Table \ref{abundtable}. Zn was detected in three of the sub-DLA absorbers
in our sample, and for the rest of the systems we place 3$\sigma$ upper limits on the Zn abundance. All of the absorbers for which Zn was detected, were found to
be metal-rich ([Zn/H] = $-0.02$ for Q1039-2719 at z$_{abs}$ = 2.139; $>-0.06$ for Q1311-0120 at z$_{abs}$ = 1.762 and $+0.25$ for Q2123-0050 at z$_{abs}$ = 2.058).
These absorbers are among the most metal-rich sub-DLAs at $z\ga1$ and are the only near-solar or super-solar metallicity sub-DLA QSO absorbers at $z\ga2$. The Zn
abundance upper limits for Q1103-2645 (z$_{abs}$ = 1.839) and Q1551+0908 (z$_{abs}$ = 2.320) place their metallicities at $<-0.82$ and $<-0.95$, respectively.
Sulphur was detected in these two systems and their metallicities based on S abundances are $-0.82$ (for Q1103-2645) and $-0.46$ (for Q1551+0908).

Table \ref{abundtable} also lists the abundance ratios of various elements along with the corresponding solar values from \citet{Lodd03}. In addition to the [Zn/Fe]
ratio, often used as an indicator of dust depletion, [S/Zn], [Si/Fe], [Cr/Fe] and [Mn/Fe] are listed. As seen from the values listed in Table \ref{abundtable}, systems
with relatively high metallicities show relatively higher dust depletion which agrees with trends found in earlier investigations. We also find evidence of
$\alpha$-enhancement, based on the [S/Zn] ratio, in two of the absorbers (toward Q1551+0908 and Q2123-0050) in our sample. Table \ref{abundtable} also lists column
density ratios between elements in different ionisation stages, which may provide information about ionisation in these systems.

\begin{table*}
\setlength{\extrarowheight}{1.75pt}
\footnotesize
\caption{Observed values of relative abundances and abundance ratios for the systems in this sample. The solar value of the ratios are given in the first row.}
\label{abundtable}
\begin{tabular*}{\textwidth}{cccccccccc}

\hline\hline
QSO		&	z$_{abs}$	&	log N$_{\rm H \ I}$	&		[Zn/H]		&		[S/H]		&		[Fe/H]		&		[S/Zn]		&		[Zn/Fe]		&		[Si/Fe]		&		[Cr/Fe]		\\
\hline																			
\vspace{1.25mm}
log (X/Y)$_{\sun}$&			&				&		$-$7.37		&		$-$4.81 	&	 	$-$4.53		&		$+$2.56		&	 	$-$2.84		&		$+$0.07		&	 	$-$1.82		\\
																			
Q1039-2719	&	2.139		&	19.55$\pm$0.15		&	$-$0.02$\pm$0.17	&	$+$0.02$\pm$0.15	&	$-$0.30$\pm$0.16	&	$+$0.04$\pm$0.08	&	$+$0.28$\pm$0.08	&	$+$0.51$\pm$0.05	&	$+$0.17$\pm$0.05	\\
Q1103-2645	&	1.839		&	19.52$\pm$0.04		&	$<$$-$0.82		&	$-$0.82$\pm$0.19	&	$-$1.45$\pm$0.04	&	$>$$+$0.01		&	$<$$+$0.62		&	$+$0.46$\pm$0.03	&	$<$$+$0.19		\\
Q1311-0120	&	1.762		&	20.00$\pm$0.08		&	$>$$-$0.06		&		...		&	$-$1.24$\pm$0.09	&		...		&	$>$$+$1.18		&	$>$$+$0.08		&	$+$0.53$\pm$0.13	\\
Q1551+0908	&	2.320		&	19.70$\pm$0.05		&	$<$$-$0.95		&	$-$0.46$\pm$0.10	&	$-$1.61$\pm$0.07	&	$>$$+$0.49		&	$<$$+$0.66		&	$+$0.28$\pm$0.06	&	$<$$+$0.41		\\
\vspace{0.75mm}
Q2123-0050	&	2.058		&	19.35$\pm$0.10		&	$+$0.25$\pm$0.12	&	$+$0.51$\pm$0.10	&	$-$0.73$\pm$0.10	&	$+$0.26$\pm$0.06	&	$+$0.98$\pm$0.06	&	$+$0.73$\pm$0.06	&	$<$$-$0.38		\\
																			
\hline																			
QSO		&	z$_{abs}$	&	log N$_{\rm H \ I}$	&	 	[Mn/Fe]    	&	Al III/Al II$^{a}$ 	&	 Fe II/Al III$^{a}$	&	Mg II/Al III$^{a}$	&	Mg II/Mg I$^{a}$	&	 Si III/Si II$^{a}$	&	 Si IV/Si II$^{a}$	\\
\hline																			
\vspace{1.25mm}
log (X/Y)$_{\sun}$&			&				&	 	$-$1.97		&				&				&				&				&				&				\\
																			
Q1039-2719	&	2.139		&	19.55$\pm$0.15		&	$-$0.27$\pm$0.07	&	$<$$-$0.35		&	$+$1.20$\pm$0.04	&		$>$$+$1.97	&		$>$$+$2.48	&	$>$$-$1.01		&	$>$$-$0.81		\\
Q1103-2645	&	1.839		&	19.52$\pm$0.04		&	$<$$+$0.89		&		...		&	$+$0.90$\pm$0.06	&		$>$$+$1.44	&		$>$$+$2.22	&	$>$$+$0.05		&	$-$0.23$\pm$0.02	\\
Q1311-0120	&	1.762		&	20.00$\pm$0.08		&	$<$$-$0.61		&	$<$$-$1.05		&	$>$$+$2.38		&		$>$$+$2.70	&		$>$$+$2.28	&		...		&		...		\\
Q1551+0908	&	2.320		&	19.70$\pm$0.05		&	$<$$-$0.19		&	$-$0.41$\pm$0.14	&	$+$1.42$\pm$0.14	&			...	&			...	&	$>$$-$0.23		&	$-$0.58$\pm$0.04	\\
\vspace{0.75mm}
Q2123-0050	&	2.058		&	19.35$\pm$0.10		&	$-$0.01$\pm$0.08	&	$<$$-$0.01		&	$+$0.64$\pm$0.07	&		$>$$+$0.82	&		$>$$+$1.53	&	$>$$-$0.73		&	$>$$-$0.89		\\
\hline

\end{tabular*}
$^{a}$Ratio of column densities.\\
\end{table*}

\subsection{Photoionisation Modelling and Ionisation Corrections}

The gas in the high H I column density absorbers is usually expected to be largely neutral due to the self-shielding of photons with $h\nu>13.6$ eV. Zn and S
in these systems are expected to be predominantly singly ionised. Consequently, the metallicities reported for such high $N_{\rm H \ I}$ absorbers are estimated
from $N_{\rm Zn II}$/$N_{\rm H I}$ or $N_{\rm S II}$/$N_{\rm H I}$ ratios. For absorbers with lower $N_{\rm H \ I}$, such estimates may not be correct if they have
non-negligible contributions from higher ionisation stages. Several studies investigating the effect of ionisation in DLAs (e.g., \citealt{Howk99,Vlad01,Pro02}) have
found that in most cases the ionisation correction factor, defined here as
$$\epsilon=[X/H]_{total}-[X^{+}/H^{0}],$$
where $[X/H]_{total}$ include contributions from all ionisation stages, is $\la0.2$ dex for most elements. Sub-DLA systems, by virtue of lower H I in them, might
be expected to show higher level of ionisation. However, it has previously been shown that the ionisation corrections are, in general, small for the sub-DLA systems
as well (e.g., \citealt{DZ03,Mei07,Mei08}).

To estimate the effect of ionisation on the sub-DLA abundances presented here, we carried out photoionisation modelling of these systems using version 13.01 of the
CLOUDY photoionisation code (\citealt{Fer13}). The models were generated assuming that the ionising radiation incident on the gas cloud is a combination of extragalactic
UV background and a radiation field produced by O/B type stars. The extragalactic UV background is adopted from \citet{HM96} and \citet{MHR99}, evaluated at the redshift
of the absorber. The O/B type stellar radiation field is based on a Kurucz model stellar spectrum for a temperature of 30,000 K. These radiation fields were mixed in
equal parts to generate the incident radiation field. It has been suggested that the contribution from local sources to the ionisation of DLA systems may not be negligible
in comparison with the background ionising radiation (\citealt{Sch06}). In addition, we also include the cosmic microwave background at the appropriate redshift of the
absorber, and the cosmic ray background in our simulation. We note however, that radiation from local shocks originating from white dwarfs compact binary systems or
supernovae was not included in our models. For each of our absorbers, grids of photoionisation models were produced by varying the ionisation parameter, defined as
$$ U=\frac{n_{\gamma}}{n_{H}}=\frac{\Phi_{912}}{cn_{H}}$$
(where $\Phi_{912}$ is the flux of radiation with h$\nu$ $>$ 13.6 eV), from $10^{-6}$ to 1. The models assumed the solar abundance pattern for the absorbers and were
tailored to match the observed N$_{\rm H \ I}$ and the observed metallicity based on N$_{\rm Zn \ II}$. Column density ratios between various ions resulting from these
grids of simulation were then compared with the observed values (see Table \ref{abundtable}) to constrain the ionisation parameter and derive the ionisation correction
values. We note, however, that ionisation in the gas depends strongly on the shape of the ionising spectrum and our assumption for the incident spectrum is one among
many possibilities. Given the assumptions described above, we can only arrive at some general conclusions regarding the strength of ionisation in the gas.

\begin{figure}

\begin{center}

\includegraphics[angle=0, width=\linewidth]{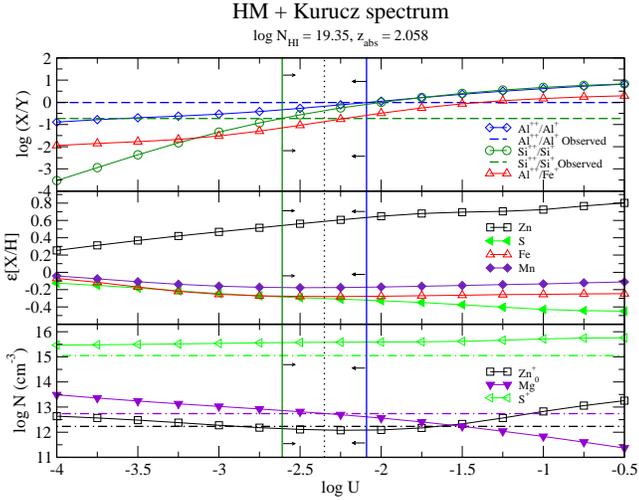}

\end{center}

\caption[Fig10]{Results of the photoionisation simulations for the sub-DLA toward Q2123-0050. The top panel shows the simulated logarithmic column density ratios of
Al$^{++}$/Al$^{+}$, Si$^{++}$/Si$^{+}$ and Al$^{++}$/Fe$^{+}$, plotted vs. the ionisation parameter. The observed upper limit for Al$^{++}$/Al$^{+}$ and the lower
limit for Si$^{++}$/Si$^{+}$ are also plotted in the same panel. The lower and upper limits on log U, determined by comparing the simulated and observed data, are
represented by the vertical solid green and blue lines, respectively. The vertical dotted line represents the mean of these limits. The panel in the middle shows the
ionisation correction factors for Zn, S, Mn and Fe abundances in dex. The bottom panel shows column density predictions from a grid of models with the corrected
metallicity incorporated in them. The comparison of the predictions with the observed column densities of Zn$^{+}$ and Mg$^{0}$, also shown in the bottom panel using
horizontal dot-dashed black and purple lines, respectively, suggests that the adopted ionisation correction to metallicity is fairly reasonable. \label{ionisationfig}}

\end{figure}

With log N$_{\rm H \ I}$ = 19.35, the sub-DLA in the spectrum of Q2123-0050 is the lowest N$_{\rm H \ I}$ system in our sample. The observed ratios of column densities in
higher ionisation stages to those in the lower ionisation stages are relatively high in this system, suggesting significant ionisation in the absorbing gas. Column density
ratios of the adjacent ions of the same element are more reliable observational constraints than the ratios involving different elements as the latter may be affected by
differential depletion or intrinsic nucleosynthetic differences. Al and Si were the elements detected in this system with multiple ionisation stages. We used the observed
lower limit of the N$_{Si^{++}}$ to N$_{Si^{+}}$ ratio to obtain a lower limit on the ionization parameter at log U $>$ -2.6. Furthermore, the observed upper limit on
N$_{Al^{++}}$/N$_{Al^{+}}$ implies log U $<$ -2.1. These results suggest that the observations underestimate the metallicity significantly as the ionisation correction for
[Zn/H] ranges between +0.54 dex to +0.63 dex. We adopt the correction to metallicity to be +0.59 dex derived for log U = -2.35, the mean value of the ionisation parameter
range described above. The corrections for [Fe/H] and [Mn/H] are derived to be -0.28 dex and -0.18 dex, respectively, suggesting a corrected value of +0.09 dex for [Mn/Fe].
The suggestion that the true depletion is much higher than observed (based on Zn II and Fe II) in a significantly ionised system (\citealt{Mei08}) seems to be true for this
system as the corrected [Zn/Fe] is $\sim$ +0.9 dex higher than the observed [Zn/Fe] = +0.98 dex. Figure \ref{ionisationfig} describes ionisation modelling results for this
system.

The observed limits on N$_{Al^{++}}$/N$_{Al^{+}}$ and N$_{Si^{++}}$/N$_{Si^{+}}$ in the log N$_{\rm H \ I}$ = 19.55 absorber toward Q1039-2719 suggest -3.1 $<$ log U $<$ -2.7.
This implies a correction for [Zn/H] between +0.45 dex and +0.51 dex. The ionisation corrections for [Zn/H], [S/H], [Mn/Fe] and [Zn/Fe], derived at the mean log U = -2.9, are
+0.48 dex, -0.20 dex, +0.08 dex and +0.67 dex, respectively.

Adjacent-ion column density ratios in the log N$_{\rm H \ I}$ = 19.52 absorber toward Q1103-2645 also suggest moderate ionisation correction to the observed abundances. The
observed lower limit on the N$_{Si^{++}}$ to N$_{Si^{+}}$ ratio allowed us to place a lower limit on the ionisation parameter at log U $>$ -3. As the Al II line was not detected
in this system, we used the N$_{Al^{++}}$/N$_{Fe^{+}}$ ratio to further constrain the ionisation parameter at log U = -2.6. The predicted correction for [S/H] was found to be
-0.31 dex. Mn and Fe abundances were only mildly affected by ionisation as shown by the estimated correction factors of -0.10 dex and -0.13 dex for [Mn/H] and [Fe/H], respectively.
We note that using the Al$^{++}$/Fe$^{+}$ ratio to estimate the ionisation parameter may introduce uncertainties due to differential depletion or nucleosynthetic differences
between the elements (see \citealt{Mei07} for a more detailed discussion on the use of adjacent ion ratios in photoionisation modelling).

The models for the absorbers toward Q1551+0908 and Q1311-0120 suggest little effect of ionisation on the observed abundances. For the log N$_{\rm H \ I}$ = 19.70 system in
the spectrum of Q1551+0908, the observed value of N$_{Al^{++}}$/N$_{Al^{+}}$ = -0.41 suggests corrections of only -0.16 dex and -0.10 dex for [S/H] and [Fe/H], respectively.
With log N$_{\rm H \ I}$ = 20.00, the sub-DLA toward Q1311-0120 is the highest N$_{\rm H \ I}$ sub-DLA in our sample and is found to be the least ionised. The limits on the
column density ratios between Al$^{++}$/Al$^{+}$ and Si$^{++}$/Si$^{+}$ constrain the ionisation parameter between -4.7 dex and -4.3 dex, limiting the correction for [Zn/H]
between +0.10 dex and +0.18 dex (+0.14 dex at the mean ionisation parameter of log U = -4.5). The ionisation corrections for [Mn/H] and [Fe/H] were found to be negligibly
small.

\subsection{Metallicity Evolution}

\begin{figure}

\begin{center}

\includegraphics[angle=0, width=\linewidth]{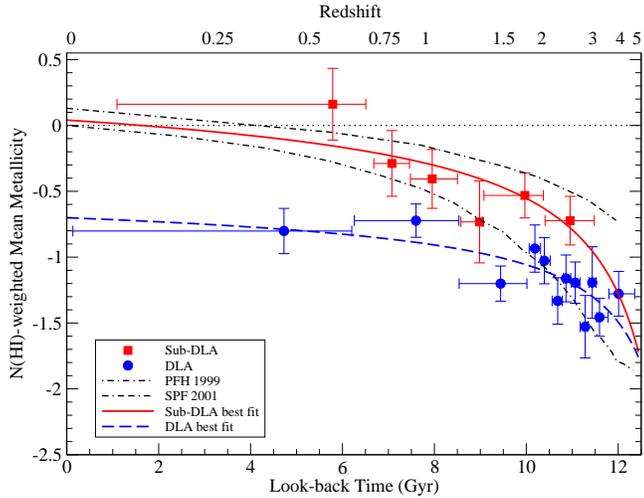}

\end{center}

\caption[Fig11]{ $N({\rm H \, I})$-weighted mean metallicity vs. look-back time relation for 195 DLAs and 68 sub-DLAs with Zn or S measurements. Filled
circles show 12 bins with 16 or 17 DLAs each. Squares denote 6 bins with 11 or 12 sub-DLAs each. Horizontal bars denote ranges in look-back times covered
by each bin. Vertical errorbars denote 1 $\sigma$ uncertainties. The bold solid and dashed curves show the best fits obtained from linear regression of
the metallicity vs. redshift data for sub-DLAs and DLAs, respectively. The light dot-dashed and dot-double-dashed curves show, respectively, the mean
metallicity in the models of \citet{Pei99} and \citet{Somm01}. Sub-DLAs appear to be more metal-rich and faster-evolving than DLAs, at all redshifts where
both DLA and sub-DLA metallicity data exist ($z\la3$). \label{metalevolfig}}

\end{figure}

We examine metallicity evolution in sub-DLAs and DLAs, by combining our data with those from the literature (\citealt{Aker05,Bat12,Boi98,Cent03,DLV00,DZ03,DZ09,EL01,Fyn11,Ge01,
Kh04,Kul99,Kul05,Led06,Lop99,Lop02,LE03,Lu95,Lu96,Mei06,Mei07,Mei08,Mei09,MY92,Mey95,Mol00,Nest08,Not08,Per02,Per06a,Per06b,Per08,Petit00,Pet94,Pet97,Pet99,Pet00,PW98,PW99,Pro01,
Pro02,Pro03a,Pro03c,Raf12,Rao05,SP01}). \citet{Raf12} presented metallicity vs. redshift relation for a larger DLA sample (242 systems) but many of their metallicity measurements
come from Si and Fe, elements prone to depletion. For our analysis, we prefer not to use Si or Fe, so as to avoid the ambiguity in estimating dust depletion corrections. Instead,
we use measurements of Zn or S (in cases where Zn was not detected), since these nearly undepleted elements provide the most direct gas-phase metallicity estimates. For systems
with no detection of Zn and S, upper limits on Zn have been used and were treated with survival analysis. N$_{\rm H I}$-weighted mean metallicity versus look-back time relations
for DLAs and sub-DLAs were determined using the procedures described in \citet{KF02}. Figure \ref{metalevolfig} shows the relations for 195 DLAs and 68 sub-DLAs in the current
sample. The DLA and sub-DLA sample are divided into 12 and 6 bins, respectively. The DLA bins contain 16 or 17 systems each, while the sub-DLA bins contain 11 or 12 systems each.

Consistent with the findings from previous studies, the current sample shows the DLAs to be generally metal poor at all redshifts probed. The sub-DLA global mean metallicity
appears to be higher than that of DLAs at all redshifts for which both DLA and sub-DLA observations are available ($0\la z\la3$). We note that although few metal rich DLAs have
indeed been observed (e.g., \citealt{Fyn11,Kh04,Nest08,Per06b}), their fraction is much lower than that of the metal rich sub-DLAs. The data also show evidence for only a weak
redshift evolution in the metallicity of DLAs. The bold solid and dashed curves in Fig. \ref{metalevolfig} show the best linear-regression fits to the $N_{\rm H I}$-weighted mean
metallicity vs. redshift data for sub-DLAs and DLAs, respectively. The linear regression estimates of the intercepts, $0.04\pm0.23$ for sub-DLAs and $-0.70\pm0.11$ for DLAs, differ
at 2.9 $\sigma$ level. The slope of the fit is estimated to be $-0.32\pm0.13$ for sub-DLAs, marginally higher than the slope, $-0.19\pm0.05$, for DLAs. It is necessary to increase
the sub-DLA sample size and to expand the sub-DLA sample at $z > 3$ to better constrain whether or not sub-DLAs evolve faster than DLAs.

Figure \ref{metalevolfig} also shows the comparison of the observations with theoretical model predictions for evolution of global interstellar metallicity. The mean interstellar
metallicity from the chemical evolution model of \citet{Pei99} is shown using the light dot-dashed curve (PFH 1999). This model calculates the coupled global evolution of stellar,
gaseous, and metal contents of galaxies by incorporating the optimum fit for the cosmic infrared background intensity and observational constraints derived from optical galaxy surveys
and the comoving H I density inferred from DLA data. The light dot-double-dashed curve (SPF 2001) represents the mean metallicity evolution of interstellar cold gas predicted by a
semi-analytic model of galaxy formation in the cold dark matter merging hierarchy by \citet{Somm01}. This model assumes a constant-efficiency quiescent star formation in addition to
starbursts triggered by galaxy mergers. It is evident from Fig. \ref{metalevolfig}, that the metallicity evolution in sub-DLAs is consistent with the chemical evolution models over
most of the redshift range probed so far, and especially at low redshifts, reaching solar level at $z=0$. The sub-DLA trend bears a closer resemblance with the merger driven `collisional
starburst model' by \citet{Somm01}. On the other hand, the DLA data are in poor agreement with the model predictions and DLA metallicity reaches only $\sim$ 1/5th of the solar value
at $z=0$. The DLA trend becomes consistent with PFH 1999 only at $z\ga2$. Some recent studies (e.g., \citealt{DO07}) predict a low DLA metallicity at  $z=0$, but do not correctly predict
the higher redshift DLA metallicities. The difference in the metallicity evolution trends in DLAs and sub-DLAs may suggest that the galaxies traced by these absorbers follow separate
evolutionary tracks established as early as $\sim$2 Gyrs after the Big Bang. However, given the small difference between the slopes of the trends, the observed difference can extend
further back in time. Sub-DLA data at redshifts higher than 3 are essential to provide further constraints on the epoch of establishment of these distinct evolutionary tracks.

Comparing the metallicities for DLAs and sub-DLAs with those for galaxies detected in emission can provide clues to the understanding of the nature of the absorbing galaxies. It is
well-known that galaxies detected in emission show a correlation between their stellar mass and the gas metallicity (e.g., \citealt{Trem04,Erb06}). Furthermore, the mass-metallicity
relation is found to evolve with redshift. \citet{Mai08} found that for star forming galaxies at $M_{*} \sim 10^{10} \, M_{\odot}$, the metallicity at $z \sim 2.2$ is lower by a factor
of about 2.5 with respect to that at $z \sim 0$. The drop is less steep for more massive galaxies, indicating that the latter got enriched at earlier epochs, consistent with the
mass-downsizing scenario. Comparing Fig. 9 of \citet{Mai08} with our Fig. \ref{metalevolfig}, the sub-DLA trend seems to resemble that for star forming galaxies with $M_{*} \sim 10^{10}
\, M_{\odot}$. The trend for DLAs, however, does not resemble any of the  trends found by \citet{Mai08} for star forming galaxies with $9 < {\rm log} M_{*}/M_{\odot} < 11$, suggesting
that DLA host galaxies have not undergone much star formation and chemical enrichment even by the current epoch. This is consistent with the observed agreement of DLA metallicity
distribution with that for the Milky Way halo stars, suggesting that most DLAs are not representative of the disks of Milky Way-type galaxies (e.g., \citealt{Pet04}).

\subsection{[Mn/Fe]-Metallicity Correlation}

\begin{figure}

\begin{center}

\includegraphics[angle=-90, width=\linewidth]{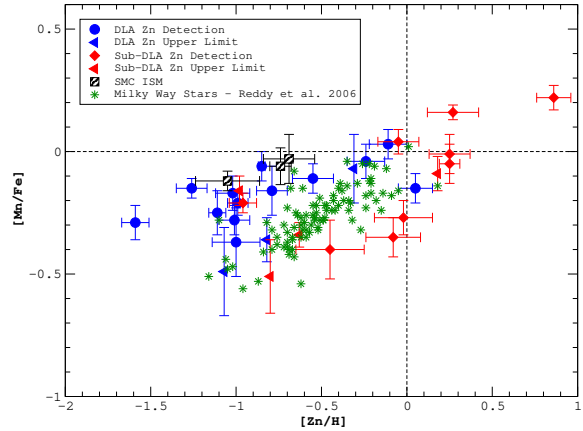}

\end{center}

\caption[Fig12]{[Mn/Fe] vs. [Zn/H] for the sub-DLAs from this sample, as well as for sub-DLAs and DLAs from the literature. Milky Way stellar abundance
data from \citet{Red06} are shown overplotted. Also shown are the interstellar abundance data for SMC from \citet{Wel01}.\label{mnfefig}}

\end{figure}

The condensation temperatures of Mn and Fe being similar, the abundance ratio between these two elements is expected to be primarily governed by
differences in their nucleosynthesis. The Mn abundance shows a strong metallicity dependence in Milky Way stars. [Mn/Fe] is also found to be
correlated with [Fe/H] in the sense that [Mn/Fe] increases with increasing [Fe/H] (e.g., \citealt{Nissen00, McW03, Grat04}). A similar trend between
[Mn/Fe] and [Zn/H] is also seen to be present in DLAs and sub-DLAs (e.g., \citealt{Mei09}). In Figure \ref{mnfefig}, we plot [Mn/Fe] versus [Zn/H]
for the absorbers in this sample, along with the data for DLAs and sub-DLAs taken from the literature. Data from \citet{Red06} for Milky Way stars
and the interstellar abundance data for SMC from \citet{Wel01} are also shown overlayed on the same plot. The trend of increasing [Mn/Fe] with increasing
[Zn/H], seen in the Milky Way stars, is clearly present in the absorber galaxies as well. Kendall's $\tau$ for the complete absorber sample (DLA +
Sub-DLA) was determined to be $\tau$ = 0.724 with the probability of no correlation being 0.002. A Spearman rank correlation test gave the correlation
coefficient $\rho$ = 0.521 with the probability of no correlation of 0.006. Although the absorber sample shows a general correlation, the dispersion
in the absorber data is larger compared to the stellar sample from \citet{Red06}. The fact that galaxies detected through absorption represent various
morphological types is likely to cause this dispersion with additional contribution from differential depletion onto dust grains between Mn and Fe.
Kendall's $\tau$ for the DLA sample alone was determined to be $\tau$ = 0.917 (with a probability of no correlation being 0.006), while $\tau$ =
0.872 for the sub-DLAs with a probability of obtaining this value by chance being 0.026. There seems to be evidence for different [Mn/Fe] versus
[Zn/H] trends between DLAs and sub-DLAs. While the DLA measurements are similar to the interstellar abundance data from the SMC, the sub-DLA data
bear resemblance with the Mn and Fe abundance pattern seen in the Galactic bulge stars (see e.g, \citealt{McW03}). The linear regression slopes for
the [Mn/Fe] vs. [Zn/H] data, being $0.12\pm0.04$ and $0.27\pm0.03$ for DLAs and sub-DLAs, respectively, differ at $\, \sim \, {\rm 3}\sigma$ level.
However, larger samples are needed to confirm this difference. A difference in the [Mn/Fe] vs. [Zn/H] relations for DLAs and sub-DLAs may suggest
a difference in the stellar populations in these two classes of absorbers.

\subsection{Velocity Dispersion-Metallicity Relationship}

Based on a sample of star-forming galaxies at z$\sim$0.1, \citet{Trem04} found a correlation between stellar mass and gas-phase metallicity for
these galaxies. Similar mass-metallicity relations have been suggested by \citet{S05} for $0.4 < z < 1.0$ galaxies selected from the Gemini Deep
Deep Survey and the Canada-France Redshift Survey and by \citet{Erb06} for UV-selected star forming galaxies at $z \sim 2.3$. \citet{Nest03} and
\citet{Turn05} noticed a correlation between the Mg II $\lambda$ 2796 equivalent width and the metallicity for strong Mg II absorbers at $1 \la
z \la 2$. The possible existence of a mass metallicity relationship for DLA absorbers, assuming the velocity width of optically thin lines to be
proportional to the mass, has recently been put into evidence \citep{Per03a, Led06}. As the velocity width of the low-ionisation absorption lines
potentially probes the depth of the underlying gravitational potential well of the DLA systems, this quantity can be used as a proxy for the stellar
mass of these systems, which has been difficult to measure. \citet{Bou06}, however, find an anti-correlation between the Mg II equivalent width and
the estimated halo mass based upon an indirect mass indicator. Also, \citet{Zwa08} show that the velocity width and mass do not correlate well in
local analogues of DLAs.

\begin{table}
\center
\footnotesize
\caption{Velocity width values for the absorbers in this sample.}
\label{kinematicstable}
\begin{tabular}{ccccc}

\hline\hline
QSO		&	z$_{abs}$	&		[Zn/H]		&	$\Delta {\rm v}_{\rm 90}$	&		Selected	\\
		&			&				&		\kms			&	transition line		\\
\hline							
Q1039-2719	&	2.139		&	$-$0.02$\pm$0.17	&			70		&	Fe II $\lambda$ 2374	\\
Q1103-2645	&	1.839		&	$<$$-$0.82		&			81		&	Fe II $\lambda$ 2600	\\
Q1311-0120	&	1.762		&	$>$$-$0.04		&			152		&	Fe II $\lambda$ 2374	\\
Q1551+0908	&	2.320		&	$<$$-$0.95		&			32		&	Fe II $\lambda$ 2344	\\
Q2123-0050	&	2.058		&	$+$0.25$\pm$0.12	&			321		&	Fe II $\lambda$ 2344	\\
\hline

\end{tabular}
\end{table}

To investigate the velocity width-metallicity relation in sub-DLAs, we measured the velocity width values for the systems in our sample following the
analysis of \citet{WP98}. The velocity width for a system was measured using an absorption profile (in velocity space) from a low-ion transition seen in
the system. High-ionisation lines are not suitable for this analysis as their velocity widths are likely to be dominated by large scale thermal motions
in the gas. The measurement method involved the conversion of the low-ion transition profile, I$_{obs}({\rm v})$, into the corresponding apparent optical
depth profile, ${\rm \tau}({\rm v})_{a}$, through the following relation
$$\tau(\rm v)_{a}=ln[{\rm I}_{0}(\rm v)/{\rm I}_{obs}(\rm v)],$$
where I$_{0}(\rm v)$ represents the continuum level, and I$_{obs}(\rm v)$ is the observed intensity of the normalised transition profile in velocity space.
The apparent optical depth was then integrated over the entire line profile to yield ${\rm \tau}_{int}$, the total optical depth within the absorption profile.
Finally, the velocity width was determined as $\Delta {\rm v}_{90} = [{\rm v}(95\%) - {\rm v}(5\%)]$, where ${\rm v}(95\%)$  and ${\rm v}(5\%)$ define the
velocity range within which 90$\%$ of ${\rm \tau}_{int}$ was contained.

In the case of very strong line profiles, the optical depth can not be measured accurately and the velocity width determined using such a line can be overestimated.
On the other hand, velocity width measured from a very weak line becomes highly sensitive to the continuum noise and can be underestimated, as part of the
absorbing gas can remain undetected. To select profiles which are neither strongly saturated nor too weak, we required the transitions profiles used to measure
the velocity widths to satisfy $0.1 < {\rm I}_{min}/{\rm I}_{c} < 0.6$, where ${\rm I}_{c}$ is the continuum level intensity, and ${\rm I}_{min}$ is the intensity
at the location of the strongest absorption in the line profile. After selecting a profile, we visually inspected the strongest low-ion transitions to ascertain
the velocity range over which the selected profile should be integrated to determine $\Delta {\rm v}_{90}$. Table \ref{kinematicstable} lists the velocity width
measurements from our systems along with the line profiles used.

\begin{figure}

\begin{center}

\includegraphics[angle=-90, width=\linewidth]{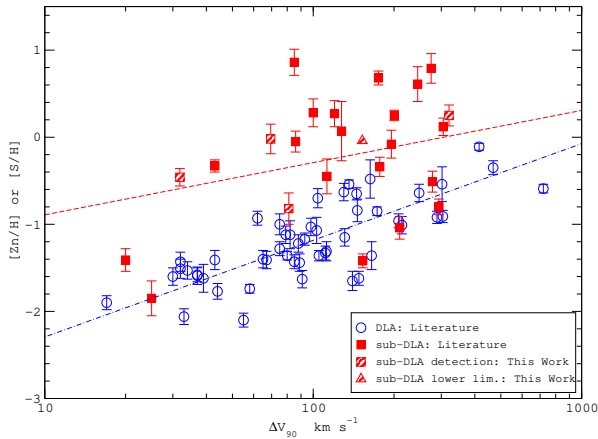}

\end{center}

\caption[Fig13]{Velocity dispersion ($\Delta V_{90}$) vs. Metallicity relations for sub-DLAs and DLAs. Linear regression fits to the sub-DLA and DLA
	        data are shown as dashed and dashed-dotted lines respectively. \label{kinematicsfig}}

\end{figure}

In Figure \ref{kinematicsfig} we plot the velocity dispersion versus metallicity based on Zn or S from our sample as well as for DLAs and sub-DLAs from
the literature. Only systems for which Zn or S was detected were plotted. A correlation between velocity width and metallicity seems to exist for DLAs,
while the sub-DLA data appear much less correlated. The sub-DLAs seem to be different from the DLAs also in terms of the mean metallicity.
A linear regression fit for the sub-DLAs gives
$$[X/H]=(0.60\pm0.07){\:}{\rm log}{\:}\Delta v_{90} -(1.49\pm0.15),$$
while a fit to the DLA data yields a slope of 1.11$\pm$0.03 and an intercept -3.40$\pm$0.06. The two slopes are different at $\, \sim \, {\rm 7}\sigma$
level.

\subsection{C II$^{*}$ Absorption and Cooling Rate}

Most of the cooling in the Milky Way's interstellar medium takes place through the fine-structure line emission of [C II] $\lambda$158 $\mu$m. This line
arises from the $^{2} P_{3/2}$ to $^{2}P_{1/2}$ transition in the ground state $2s^{2} \, 2p$ term of C II. Following \citet{Pot79}, the rate of cooling
per H atom in gas detected in absorption can be expressed as:
$$l_{c} = \frac{N_{\rm C II^{*}} h \nu_{ul} A_{ul}}{N_{\rm H I}}{\:}{\:}{\:}{\rm ergs}{\:}{\:}{\rm s^{-1}},$$
where $N_{\rm C II^{*}}$ is the column density of the C II ions in the 2P$_{3/2}$ state, $N_{\rm H I}$ is the H I column density, while $h \nu_{ul}$ and
$A_{ul}$ are the energy of the $^{2} P_{3/2}$ to $^{2}P_{1/2}$ transition and coefficient for spontaneous photon decay, respectively. UV transitions of
C~II$^{*} \, \lambda 1335.7$ and Ly$\alpha \, \lambda 1215.7$ can be used to infer N$_{\rm C II^{*}}$ and N$_{\rm H I}$, respectively, for the determination
of l$_{\rm c}$ in the interstellar medium detected in absorption.

Our data shows the presence of C~II$^{*} \,\lambda 1335.7$ in the sub-DLAs toward Q1039-2719, Q1103-2645 and Q2123-0050. However, this line is partially blended
with C~II$\lambda 1334$ in Q1039-2719 and Q2123-0050 (see Figures \ref{q1039ionsfig} and \ref{q2123ionsfig}, respectively) while for Q1103-2645, it is partially
blended with a Ly$\alpha$ forest feature. As a result, only a lower limit on N$_{\rm C II^{*}}$ could be placed for each of these absorbers. However, the
absorption profile structures of these systems suggest that the true N$_{\rm C II^{*}}$ values are unlikely to be much higher than the corresponding lower limits.
For the sub-DLA toward Q1551+0908, C~II$^{*} \,\lambda 1335.7$ was not detected and we placed a 3$\sigma$ upper limit on N$_{\rm C II^{*}}$ based on the S/N near
the line. Poor S/N in the region of the transition did not allow us an estimate of C II$^{*}$ abundance in the sub-DLA toward Q1311-0120. Table \ref{coolingratetable}
lists the N$_{\rm C II^{*}}$ and the corresponding l$_{\rm c}$ values for the sub-DLAs in this sample.

\begin{table}
\setlength{\tabcolsep}{5.6pt}
\setlength{\extrarowheight}{1.75pt}
\center
\footnotesize
\caption{Cooling rate values for the absorbers in this sample}
\label{coolingratetable}
\begin{tabular}{cccc}

\hline\hline
QSO		&	log N$_{\rm H \ I}$	&		N$_{\rm C II^{*}}$		&			l$_{\rm c}$			\\
		&		cm$^{-2}$	&			cm$^{-2}$		&		ergs s$^{-1}$ per H atom		\\
\hline							
Q1039-2719	&	19.55$\pm$0.15		&	$> 3.92 \times 10^{13}$		&		$> 3.33 \times 10^{-26}$		\\
Q1103-2645	&	19.52$\pm$0.01		&	$> 8.53 \times 10^{12}$		&		$> 7.80 \times 10^{-27}$		\\
Q1551+0908	&	19.70$\pm$0.05		&	$< 1.48 \times 10^{12}$		&		$< 8.93 \times 10^{-28}$		\\
Q2123-0050	&	19.35$\pm$0.10		&	$> 6.60 \times 10^{13}$		&		$> 8.90 \times 10^{-26}$		\\
\hline

\end{tabular}
\end{table}

\begin{figure}

\begin{center}

\includegraphics[angle=0, width=\linewidth, trim= 0 0 0 -49]{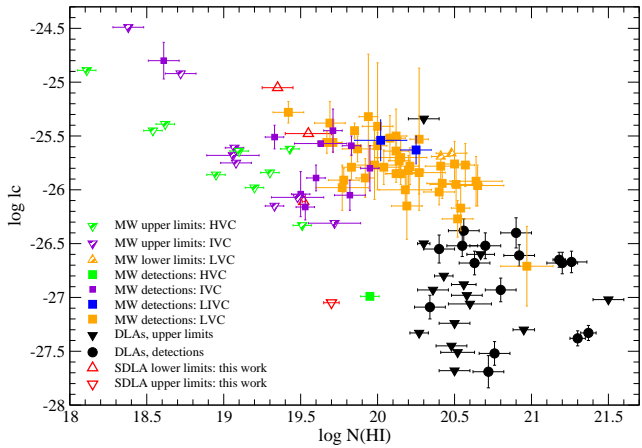}

\end{center}

\caption[Fig14]{Cooling rate estimated from C II$^{*}$ absorption plotted vs. H I column density. The open red triangles represent the sub-DLAs from our
sample. The filled black circles and triangles denote the sample of QSO DLAs in \citet{Wol03}. The filled squares and striped triangles represent the
measurements for low, intermediate, low+intermediate, and high-velocity interstellar H I clouds in the Milky Way  compiled in \citet{Leh04}.\label{coolingratefig}}

\end{figure}

The cooling rate versus H I column density data for these sub-DLAs are plotted in Figure \ref{coolingratefig}, along with the corresponding measurements for DLAs from
\citet{Wol03} and for interstellar clouds in the Milky Way adopted from \citet{Leh04}. The ISM measurements are shown separately for low, low+intermediate, intermediate,
and high-velocity clouds in the Milky Way. Although our measurements could only provide limits on the sub-DLA cooling rates, it can immediately be inferred from Fig.
\ref{coolingratefig} that, with the exception of the absorber toward Q1551+0908, these systems show higher cooling rates compared to the QSO DLAs and similar values to
those seen in the Milky Way interstellar clouds. We also note, assuming the true cooling rates lie close to the observed lower limits, that the sub-DLA showing the highest
cooling rate is also the most metal-rich absorber in our sample while the system with the least metallicity happens to show the lowest l$_{\rm c}$ value. This, combined
with the measurements on the other absorbers from our sample, points to the possibility that the cooling rate in sub-DLAs may increase with metallicity. However, a
detailed investigation of the metallicity dependence of cooling rate in sub-DLAs warrants a much larger sample with precise column density determinations.

\section{Conclusions}

In this paper, we have presented high-resolution absorption spectra of 5 sub-DLAs at $1.7<z_{abs}<2.4$. Although, to date the DLA systems have been the preferred tracer of
metallicity at high redshift, most of the absorbers observed to have solar or higher metallicity have been sub-DLAs (e.g., \citealt{Pet00,Kh04,Per06a,Pro06,Mei07,Mei08,Mei09}).
With the sub-DLA sample presented in this paper, we have found a system with ${\rm [Zn/H]}>-0.06$ dex at $z_{abs}=1.76$ and two systems with ${\rm [Zn/H]}=+0.25$ dex and ${\rm
[Zn/H]}=-0.02$ dex (+0.84 dex and +0.48 dex, respectively, after ionisation correction) at $z_{abs}>2$. These two systems are the most metal-rich sub-DLAs known so far
at $z_{abs}\ga2$. These observations suggest that metal-rich sub-DLAs appear at high redshift as well. Combining the data presented in this paper with other sub-DLA and DLA
data from the literature, we have also reported the most complete existing determination of the N$_{\rm H \ I}$-weighted mean metallicity vs. redshift relation for sub-DLAs and
DLAs. The results show that the trend of higher mean metallicity in sub-DLAs compared to DLAs, observed previously at $z<1.5$, continues to exist at least upto $z\la3$.
We also find that while metallicity evolution in DLAs does not resemble the expected mean trend for chemical enrichment in galaxies, the sub-DLA data are consistent with the chemical
evolution models at all redshifts probed so far. It is possible that most of the DLA host galaxies have not undergone much star formation even by the current epoch but the majority of
the sub-DLAs trace massive star forming galaxies. To gain additional insights into the nature of DLAs and sub-DLAs, we compared their [Mn/Fe] vs. metallicity trends and the
results suggest a difference in the stellar populations for the galaxies traced by these two classes of QSO absorbers. We also compare the velocity dispertion vs. metallicity
trends for these absorbers and find that, while metallicity correlates with velocity dispertion in DLAs, the sub-DLA data show a lower degree of correlation. Finally, we estimated
cooling rates for the sub-DLAs in our sample using the C~II$^{*} \,\lambda 1335.7$ line, and compared them with the DLA data available in the literature. The observed lower limits
suggest that metal rich QSO sub-DLAs can show higher cooling rates than those seen in the QSO DLAs.

\section*{Acknowledgments}

We are grateful to an anonymous referee for comments that have helped to improve this paper. We also thank the helpful staff of Las Campanas Observatory for their assistance during
the observing runs. D. Som and V. P. Kulkarni gratefully acknowledge support from the U.S. National Science Foundation grants AST-0908890 and AST-1108830 (PI Kulkarni).

\bsp

\label{lastpage}

\end{document}